**Crystal Anisotropy Implications on the Magneto-Optical Properties of van der Waals $FePS_3$**

*Ellenor Geraffy[1#], Kusha Sharma[1#], Shahar Zuri[1], Faris Horani[1,2], Adam K. Budniak[1], Muhamed Dawod[3], Yaron Amouyal[3], Thomas Brumme[4], Andrea Maricel León[7], Thomas Heine[4,5,6*], Rajesh Kumar[8], Doron Naveh[8], and Efrat Lifshitz[1*]*

[1]Schulich Faculty of Chemistry, Solid State Institute, Russel Berrie Nanotechnology Institute, Grand Program for Energy and the Helen Diller Quantum Center, Technion-Israel Institute of Technology, Haifa 3200003, Israel

[2]Department of Chemistry, University of Washington, Seattle, Washington 98195-1700, United States

[3]Department of Materials Science and Engineering, Technion – Israel Institute of Technology, 3200003 Haifa, Israel

[4]Faculty of Chemistry and Food Chemistry, Technical University of Dresden, 01069, Dresden, Germany

[5]Center for Advanced Systems Understanding, CASUS, HZDR, 02826 Görlitz, Germany

[6]Department of Chemistry and IBS for nanomedicine, Yonsei University, Seoul 120-749, Republic of Korea

[7]Departamento de Física, Facultad de Ciencias, Universidad de Chile, Casilla 653, Santiago, Chile

[8]Faculty of Engineering, Bar-Ilan University, Ramat-Gan Israel 52900

#Equal contribution

*Corresponding authors

Email: ssefrat@technion.ac.il; thomas.heine@tu-dresden.de

**Abstract.** Antiferromagnetic $FePS_3$ has recently gained significant interest in its potential applications in spin-related devices. Here, we show that in-plane structural anisotropy has a major impact in shaping the optical responses of $FePS_3$ single-crystals from the bulk form down to the monolayer limit. X-ray diffraction on a bulk $FePS_3$ crystal confirms a distorted $FeS_6$ octahedron causing inequivalent Fe-Fe distances and consequently resulting in a higher a/b lattice parameter ratio. Micro-photoluminescence observations on bulk and monolayer $FePS_3$ reveal four emissions: one intra-atomic *d-d* transition (band A, centered at ~1.24 eV) and three *p-d* charge transfer transitions (bands B, C, and D, centered around ~1.79 eV, ~2.3 eV, and ~2.56 eV, respectively). These bands exhibit different polarization behaviors, which persist down to the monolayer limit. Density functional theory calculations from bulk to monolayer $FePS_3$ reveal the underlying electronic structure, assign the observed emissions, and indicate why these peaks have contrasting linear and circular polarization responses. These results establish a direct structure-optics relation in $FePS_3$, highlighting the strong coupling between lattice anisotropy, electronic transitions, and symmetry-selective optical selection rules.

## 1. Introduction

Two-dimensional (2D) van der Waals magnetic materials like transition-metal phosphorus trichalcogenides have regained widespread attention for their electronic, optical, and magnetic properties in both bulk and monolayer forms,[1,2] making them promising materials for spintronic and spin-optical applications.[3,4] With the general formula $MPX_3$ (M = first-row transition metal (e.g., Fe, Mn, Ni), P = phosphorus, X = S or Se), each layer features a 2:1 arrangement of edge-sharing $[MX_6]^{2+}$ and $[P_2X_6]^{4-}$ octahedra, with the metal ions forming a honeycomb lattice.[5–10] These layers are bound by weak vdW forces, enabling facile exfoliation down to the monolayer, akin to graphene. The unique structure of $MPX_3$ materials supports antiferromagnetic (AFM) ordering of metal ions arranged in a honeycomb lattice, with spins aligning either out-of-plane (Ising-type or Heisenberg-type) or in-plane (XY-type)[5,8] in various AFM configurations, including AFM-Néel,[7] AFM-zigzag,[5,7] and AFM-stripy.[11] The magnetic order arises from exchange interactions between unpaired d-electrons, extending up to third nearest neighbors and enabling long-range coherence. Among the $MPX_3$ materials, $FePS_3$ presents a special case, possessing $Fe^{2+}$ $d^6$ electronic configuration in the octahedral field,[1,12] which bestows a strong spin–orbit coupling and a trigonal distortion of the $FeS_6$ octahedra units Consequently, $FePS_3$ hosts an Ising-type zigzag AFM ground state with an easy-axis down to monolayers, has a Néel temperature ($T_N$) just below ≈120 K, and crystallizes in a slightly distorted monoclinic *C2/m* lattice (β ≈ 107.1°), which breaks the ideal honeycomb symmetry.[13]

**Figure 1** compares the isotropic $MnS_6$ octahedron in $MnPS_3$ with the distorted $FeS_6$ octahedron, where the Fe–S bond lengths are inequivalent. Temperature-dependent X-ray diffraction (XRD) measurements with the aid of Density Functional Theory (DFT) calculations of nearest-neighbor spin-exchange interactions in the distorted $FeS_6$ honeycomb lattice, provided in the Supporting Information (SI), show that the a/b lattice-constant ratio changes markedly with temperature.[14] Thus, implying the presence of a local symmetry breaking in the honeycomb framework and corresponding shifts in nearest-neighbor (NN) distances. These in-plane distortions, in turn, can alter the material's optical response and favor a magnetic crossover from the low-energy zigzag type antiferromagnetic state to full ferromagnetism, given their close energetic proximity.[1,12,15] Olsen et al. used DFT + U to map the exchange constants up to third-nearest neighbors, finding that trigonal distortion shifts the calculated $T_N$ and strengthens the out-of-plane single-ion anisotropy (energy difference between in- and out-of-plane spins) of ~0.45 meV/Fe ion.[16] Lee et al. later verified these findings via X-ray photo-emission electron-microscopy on exfoliated flakes, revealing the single-ion anisotropy to be ~22 meV/Fe ion and attributed it to unquenched orbital moments endowed by strong spin-orbit coupling (SOC).[17] The exact orientation of this zigzag axis remains debated: some authors align it with the crystallographic a-axis,[18] whereas others place it within the a/b plane.[15,19]

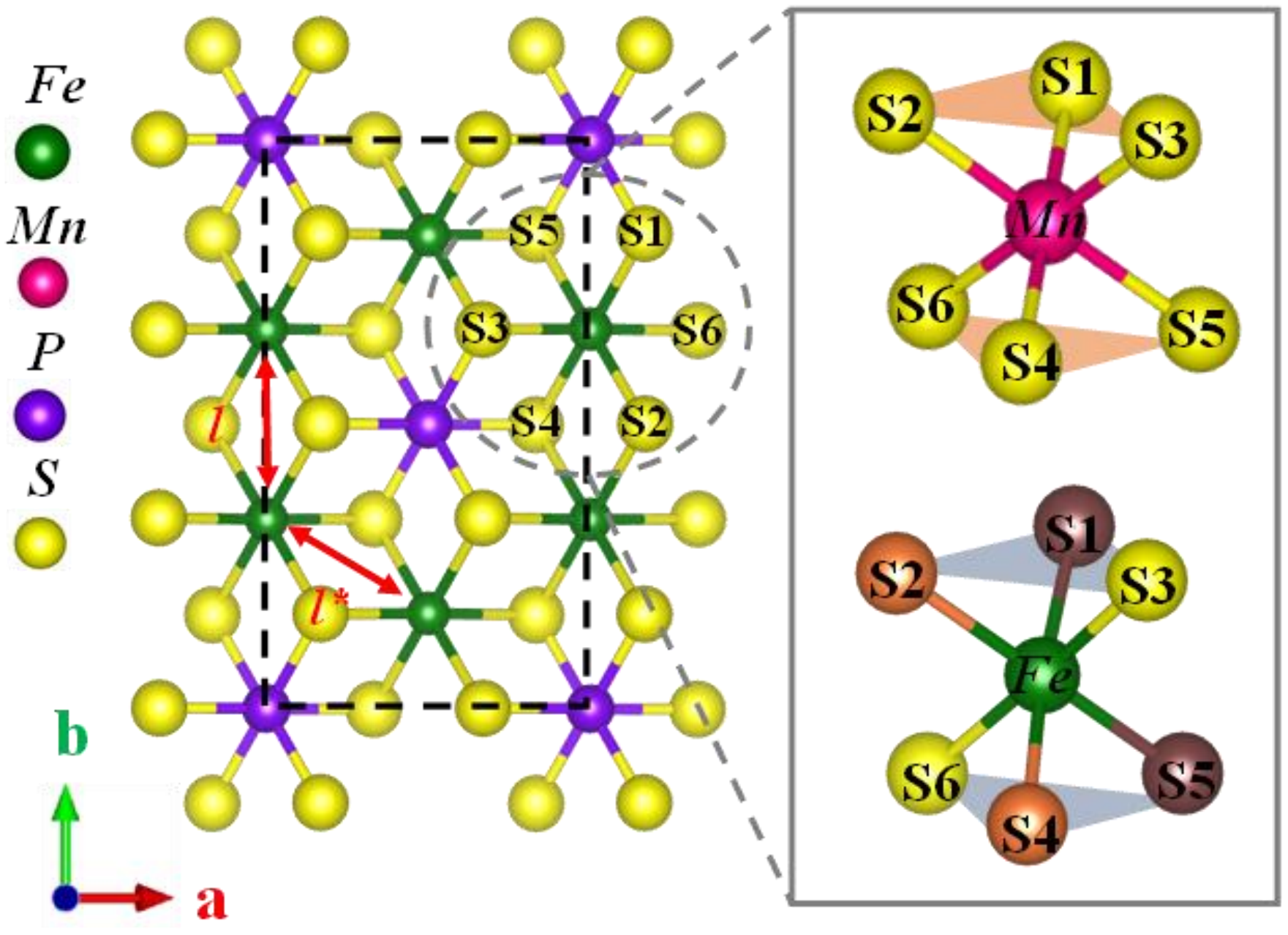

**Figure 1. Crystallographic structure of $MPX_3$ materials.** Crystal lattice representation of $MPX_3$ materials, with a close-up of an isotropic $MnS_6$ octahedra and anisotropic $FePS_6$ octahedra. Non-equivalent Fe-S bonds are marked by different colors of the S atom, highlighting a local inversion symmetry breaking in $FePS_3$. Varied Fe-Fe bond lengths are marked by l and l*.

The inherent structural anisotropy of $FePS_3$ leaves distinct fingerprints on its electronic and optical responses. Optical studies by Zhang et al.[18] showed that the polarization-resolved reflectance on exfoliated $FePS_3$ flakes below $T_N$ revealed a giant linear dichroism (LD) between light polarized parallel to the a- and b-axes. They also noted that this giant LD, although drops with the reduction in layers, persists down to the monolayer.[18] Recent trARPES studies by Nitschke et al. provided the first momentum-resolved view of how SOC and exchange anisotropy split the Fe 3d manifold above $T_N$, by identifying two distinct relaxation channels: spin-allowed and spin-forbidden d–d transitions with distinct ultrafast decay timescales, which map directly onto the anisotropic exchange.[20] In a recent μ-ARPES study, Pestka et al. analyzed the band structure of exfoliated $FePS_3$ across $T_N$, discovering three orbital-selective changes, revealing a complex interplay between orbitals and magnetic exchange driven by super-exchange interactions.[21] Ghosh et al. reported anisotropic magnetodielectric coupling in $FePS_3$, revealing that above $T_N$, the dielectric response along the c-axis shows frequency-dependent relaxations, while the in-plane response is frequency-independent. At low temperatures (below 40 K), the dielectric constant exhibits an additional anomaly distinct from the behavior near $T_N$. These contrasting trends point to a highly anisotropic spin-phonon coupling.[22] The intrinsic anisotropy of $FePS_3$ has been influential and can be harnessed to engineer a variety of heterostructures with other van der Waals materials, enabling the exploration of their combined physical effects. In one such research, Chen et al. demonstrated that when monolayer $WSe_2$ is stacked on a $FePS_3$ substrate, the interlayer exciton acquires linear polarization, proving that the low-symmetry lattice and exchange field of $FePS_3$ dominate the valley selection rules of the adjacent semiconductor.[23] Band-alignment engineering with monolayer $MoS_2$ and $MoSe_2$ likewise enhances charge transfer and photoluminescence, demonstrating the versatility of $FePS_3$ as an anisotropic material.[24]

Despite all these advances, several fundamental questions remain unsolved. This study tackles the open question of how in-plane structural anisotropy impacts the optical response of $FePS_3$ from the single-crystal bulk down to the monolayer form. Micro-photoluminescence (μPL) was conducted under both linear and circular polarization and at varying temperature ranging from 4K-120K (i.e. below and above $T_N$). These measurements revealed four emission transitions, one intra-atomic *d-d* transition (labeled band A) and three *p-d* charge transfer transitions (labelled bands B-D). Complementary single-crystal X-ray Diffraction (XRD) showed a markedly enlarged a/b lattice-parameter ratio, owing to the distorted $FeS_6$ octahedron and resulting inequivalent Fe-Fe distances (see **Figure 1**(inset) for comparison of isotropic $MnS_6$ octahedron with the distorted $FeS_6$ octahedron, highlighting the unequal Fe–S bonds). Electronic structure of $FePS_3$ from the bulk to the monolayer form was calculated using DFT which revealed the origins of these emissions and provided an explanation, for the first time, on the *p-d* charge transfer transitions' varied responses under linear and circular polarizations.

## 2. Results

Single-crystal bulk $FePS_3$ was synthesized using the chemical vapor transport (CVT) method.[13] The obtained crystals were analyzed by energy-dispersive X-ray spectroscopy (EDX), revealing a stoichiometric composition of 1:1:3 (Fe:P:S), as shown in **Figure S1** and **Table S1** of the Supplementary Information (SI). Further, thin flakes were mechanically exfoliated from bulk $FePS_3$ onto a $SiO_2$/Si substrate, yielding regions spanning from few-layer to monolayer thicknesses. These exfoliated flakes were capped with a thin h-BN flake using the dry transfer method in a glovebox to inhibit oxidative degradation. Single-crystal XRD measurements at 140 K and room temperature exposed structural anisotropy in the bulk single-crystal $FePS_3$ (see **Tables S2-S5** in Supplementary Information (SI)), which may have a significant influence on the magnetic and optical properties (see below).

To reveal the implications of the structural anisotropy in $FePS_3$ on its optical properties, various μPL measurements were conducted on bulk single crystals and their exfoliated counterparts.

The spatial resolution endowed by μPL enabled selective probing of regions with distinct thicknesses.

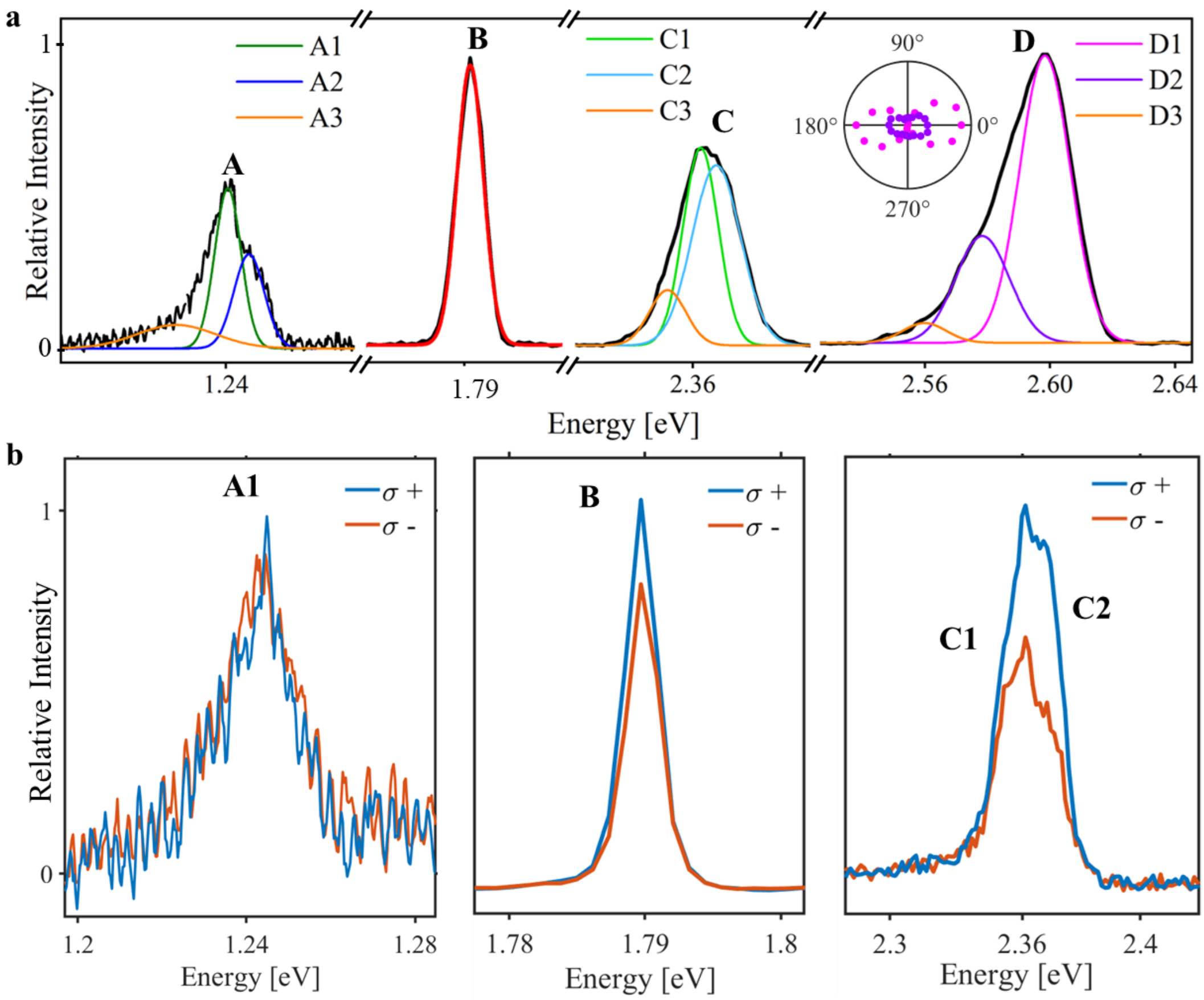

**Figure 2**. **Overview of entire emission spectrum of bulk single-crystal $FePS_3$. (a)** Four main transition bands observed under a 405 nm diode laser excitation at 4K – A (A1 and A2), B, C (C1 and C2), and D (D1 and D2). **Inset**: Polar plots of bands A and D, to assess their elliptical to linear polarization. **(b)** Emission bands A-C measured under two opposing circular polarization detections, σ+ in blue and σ- in orange.

**Figure 2(a)** shows the full μPL spectra obtained by exciting the bulk $FePS_3$ single-crystal sample above the bandgap of 1.4 eV[13] using a 405 nm (i.e., ~3 eV) diode laser at 4 K. Unless specified otherwise, all the optical measurements were carried out under these conditions. Four emission bands were revealed, centered at 1.24 eV, 1.79 eV, 2.3 eV, and 2.56 eV - designated as bands A, B, C, and D, respectively. To reveal their underlaying components, Gaussian fittings were applied to all bands, as guided by their corresponding second derivative plots (see SI, **Figure S5**). In addition, to identify whether the emission components exhibit linear (LP) or circular polarization (CP), each band was recorded under linear and circular polarizers (in the absence as well as the presence of an external magnetic field, see **Figures S7** & **S8** in SI). The representative polar plots of the LP measurements are given in the inset of panel (a). **Figure 2(b)** shows the CP measurements of bands A, B, and C under opposite helices. The degree of circular polarization (DCP) was calculated as:

$$DCP(\%) = \frac{I_{\sigma^+} - I_{\sigma^-}}{I_{\sigma^+} + I_{\sigma^-}}\ x\ 100 \qquad (1)$$

where $I_{\sigma+/\sigma-}$ is the intensity of the σ± circular polarized component.

In band A, the fittings reveal two distinct features (excluding the low-energy shoulder in orange) - peak A1 at 1.240 eV (in green color) and peak A2 at 1.247 eV (in blue). Nitschke et al. reported a similar emission composed of a weak and broad peak (similar to A2) at ~1.1 eV assigned as $^5T_{2g} \rightarrow {}^5E_g$, a spin conserving transition[20] and a lower energy shoulder (similar to A1) at ~ 0.98 eV, attributed to the to the splitting of the $^5T_{2g}$ level into the $^5E_g$ and $^5A_{1g}$ states.[25] As sub-bandgap d-d transitions (well below the reported bandgap of ~1.4 eV[13]) these features are intrinsically weak, as supported by the emission observed in **Figure 2(a)**. A lack of LP and CP behavior was observed in both A1 and A2 peaks.

Band B is composed of a single component at 1.79 eV, which lies very close to the ~1.78 eV feature reported in the literature, consistent with the $^5T_{2g} \rightarrow {}^3T_{1g}$ spin-forbidden intra-ionic d–d transition in $FePS_3$.[20] However, the sharp and intense feature of band B in **Figure 2(a)** may suggest an origin other than a spin-forbidden $Fe^{2+}$ d–d transition, which is typically weak. Prior studies report a strong linear dichroism in $FePS_3$ across the 1.4-2.0 eV window, including pronounced polarization near 1.44 eV and 1.65 eV,[18] stemming from zigzag AFM magnetic ordering. Trigonally distorted crystal fields and strong SOC in anisotropic vdW antiferromagnets like Ising $FePS_3$ and XY $NiPS_3$ create a built-in lattice strain, mediated by unequal charge distribution, leading to and define a preferred orbital axis (or magnetic easy axis). This in turn couples the optical transition dipole to the crystallographic/magnetic easy axes, resulting in pronounced linear dichroism aligned with those axes. By contrast, d–d transitions are often reported as essentially unpolarized. However, they can show appreciable LP when coherently coupled to phonons or magnons,[26,27] typically accompanied by spectral broadening. In our case, band B is sharp and intense and shows no LP. Moreover, band B shows negligible CP with DCP of ~10% that remains unchanged even upon increasing the external field to 6T (as shown in SI, **Figure S8(a)**). An explanation for the lack of polarization in band B will be provided in the discussion section. Given the sharp and narrow appearance of band B, control experiments were conducted to show that the origins of the band is not from Cr impurities (see SI, **Figure S6(a)**).

Band C (excluding the low energy shoulder) exhibits two distinct components, C1 (in light green) and C2 (in light blue), centered at 2.33 and 2.39 eV. Aruchamy et al. identified an indirect higher-energy transition at ~2.3 eV in $FePS_3$, to a transition from a valence band state into a conduction band comprised from the P–P and P–S antibonding orbitals.[28] In band C, both C1 and C2 display a CP behavior with a DCP of ~ 20.89%, and 30.43%, respectively. However, under an external magnetic field, the DCP values for C1 and C2 decrease markedly, reaching approximately 0% at 3 T and −5% at 6 T (given in SI, **Figure S9(a)**), which may be attributed to states' mixing induced by the magnetic field.

Band D (excluding the low energy shoulder which may correspond to a phonon replica) constitutes two peaks D1 and D2, identified at 2.57 eV (in magenta), and 2.59 eV (in light pink), respectively, with D2 residing very close to the 2.58 eV $^5T_{2g} \rightarrow {}^3T_{2g}$ spin forbidden transition observed and assigned by Joy et al.[29] Peaks D1 and D2 exhibit clear LP behavior, which may be a direct influence of the zigzag AFM orientation in $FePS_3$ material, as discussed above.[18] Further details on the origins of band C and D and their contrasting polarizations will be provided in the discussion section.

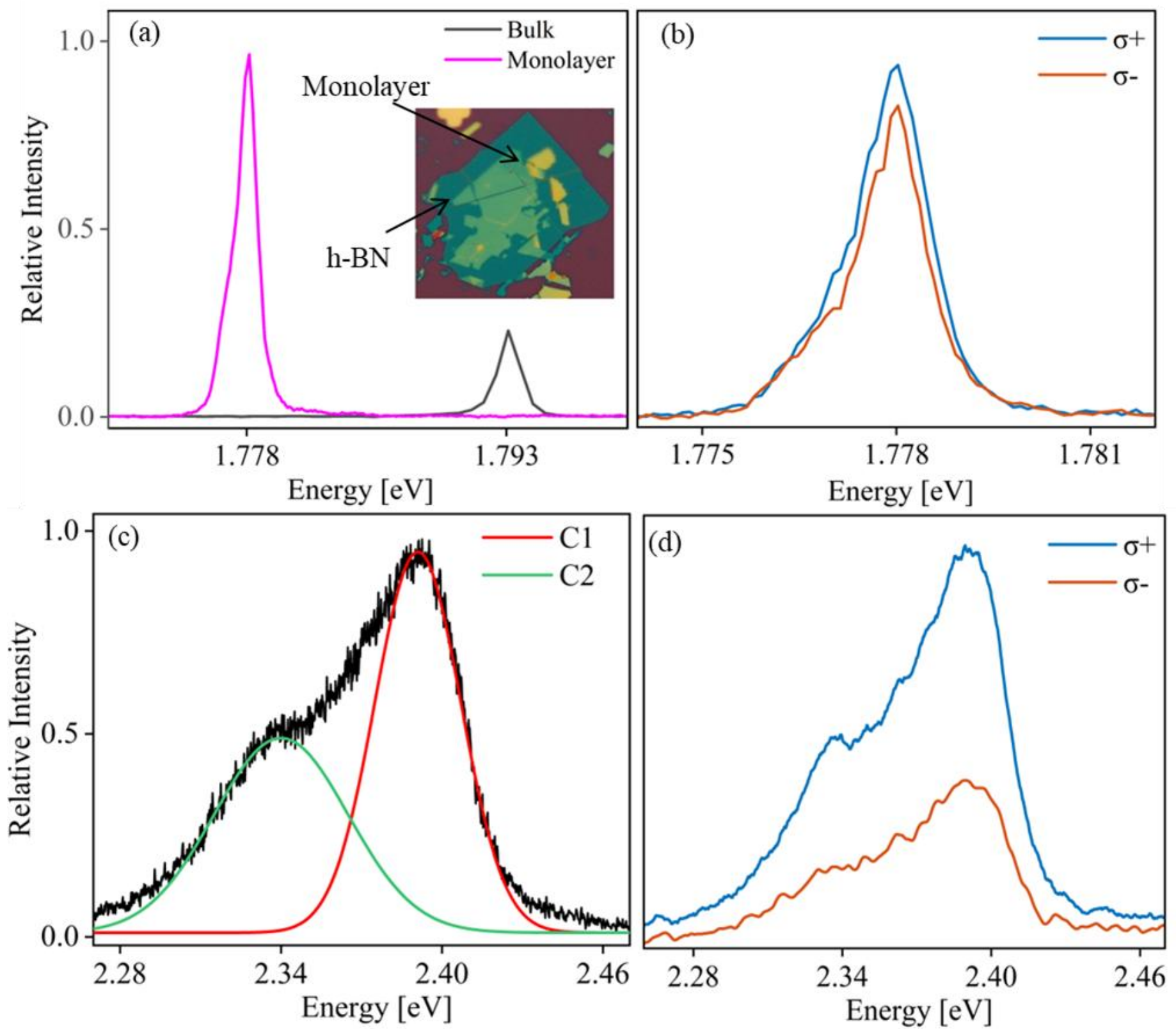

**Figure 3. Emission spectra of bands B and C in monolayer $FePS_3$. (a)** Comparison of bulk (in black) and monolayer (in pink) emission spectra in band B region. **(b)** Monolayer emission band B measured under two opposing circular polarization detections, σ+ in blue and σ- in orange **(c)** Monolayer emission spectra in band C, showing two distinctive peaks – C1 (in red) and C2 (in green) extracted using Gaussian fitting. **(d)** Monolayer emission band C measured under two opposing circular polarization detections (σ±).

Furthermore, to investigate the influence of sample thickness on the behavior of the aforementioned bands and their constituents, exfoliated samples containing few-layer regions (~20–40 layers) as well as monolayers were examined via μPL measurements at 4 K and were measured under two opposing CP detections, σ+ and σ-. **Figure 3** shows the emission spectra of bands B and C and their corresponding circular polarization responses in the monolayer limit.

Band A could not be observed in the monolayer samples of $FePS_3$. Its absence may stem from the thickness-induced band-gap shift, which moves the absorption band-edge over (or too close to) band A, thereby masking it. Further details on this phenomenon will be provided in the discussion section. **Figure 3(a)** compares the emission spectra of band B in the bulk and the monolayer limit, showing that the monolayer emission appears red-shifted by ~15 meV relative to its bulk counterpart. This red-shift may result from strain developed during the preparation of the exfoliated samples via the dry-transfer method, which is consistent with the well-documented strain sensitivity of $MPX_3$ materials.[30,31] First-principles studies for monolayer $NiPS_3$ show a strain-driven red-shift of the main optical peaks,[31] while high-pressure optical spectroscopy on $MnPS_3$ reveals the absorption edge moving to lower energy with increasing (compressive) strain, driving the optical emission to lower energy in the case of monolayered samples.[30,31] In addition, the capping with h-BN introduces dielectric screening to the monolayer $FePS_3$, which may lead to a red-shift.[32–34] The increase in intensity of band B in monolayer sample (as compared to the bulk) may originate from either (i) strong exciton binding energy, and reduced inter-layer charge transfer that results upon reducing the thickness of a bulk crystal, as shown by numerous other studies[35,36] or (ii) the capping of the flake with h-BN, which has been shown to significantly improve the PL quantum yield of the material as well as reduces the exciton annihilation rate.[37] **Figure 3(b)** presents the CP of band B in the monolayer limit, exhibiting only ~9% DCP, which remains unchanged even under the external magnetic field of 6T (see SI, **Figure S8(b)**). The absence of LP and DCP in band B in monolayer $FePS_3$ is consistent with the observation by Tan et al. in $NiPS_3$, where d-d transitions were observed to be non-polarized.[38]

**Figure 3(c)** displays band C at the monolayer limit, fitted using Gaussian fitting into two components, C1 (in red) and C2 (in light green), at 2.33 eV and 2.39 eV, respectively. At the monolayer limit, the energy difference between C1 and C2 widens, as C1 is red-shifted by ~28.5 meV and C2 blue-shifted by ~26 meV as compared to its bulk counterpart (shown in **Figure 2(a)**). **Figure 3(d)** shows CP peaks C1 and C2 in the monolayer limit, possessing a DCP of 37 % and 41%, respectively. It should be noted that in case of band C, the DCP increased as the thickness is reduced from bulk to monolayers. In addition, the DCP values for band C at the monolayer limit remain unchanged under an external magnetic field of up to 6T (SI, **Figure S9(b)**). Further explanation for why no significant change is observed from the bulk $FePS_3$ to the monolayer form in bands B and C will be provided in the discussion section.

**Figure 4(a)** shows band D in the monolayer sample, with the main peak D1 at 2.59 eV and a lower energy shoulder at 2.56 eV labelled as D2. Furthermore, polarization experiments were conducted to ascertain if the LP behavior of band D in the bulk form persisted down to the monolayer limit. The polar plot shown in the inset of **Figure 4(a)** revealed clear linear polarization behavior in peaks D1 (in blue) and D2 (in green) in the opposing directions. A color map of the entire linear polarization measurement on band C is shown in **Figure 4(b)**, where peaks D1 and D2 are clearly visible at orthogonal polarizing angles. The orthogonal LP character of D1 and D2 in the monolayer is consistent with exfoliation-induced strain effects that further perturb the anisotropic trigonal crystal field of the $FeS_6$ octahedra, thereby rotating the LP principal axes of the two components of band D in monolayer limit as compared to the bulk. [18]

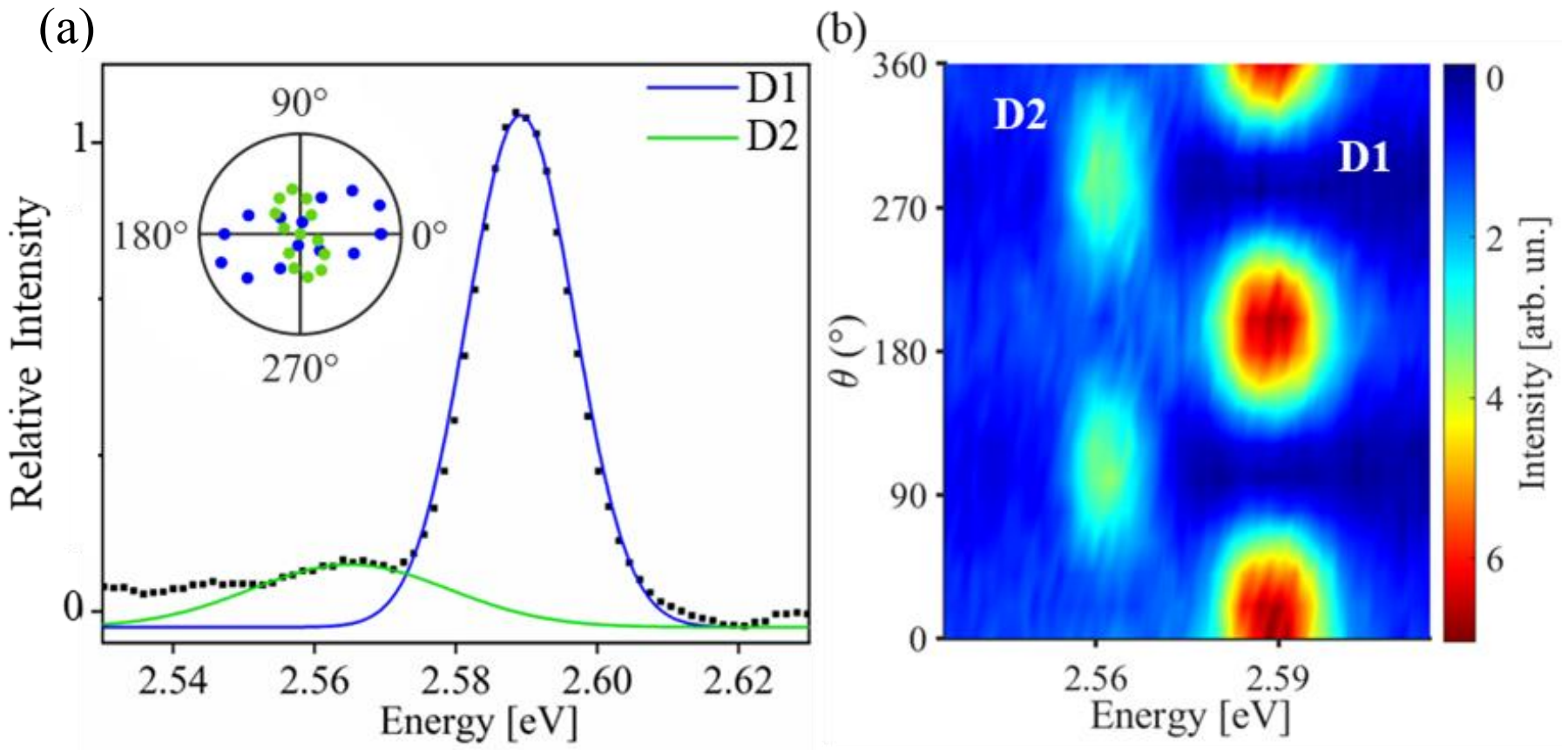

**Figure 4. Emission spectra of band D in monolayer $FePS_3$ (a)** Emission spectra of band D region where like in the bulk form, 2 peaks – D1 (in blue) and D2 (in green) were extracted from Gaussian fits. **Inset:** Polar plot of D1 and D2 where a high degree of linear polarization in opposing directions is observed. **(b)** Color map of full linear polarization measurement on the band D region in the monolayer limit, where D1 and D2 are labelled in white.

To evaluate the robustness of the zigzag AFM arrangement in $FePS_3$, temperature-dependent μPL spectra of bands A–D were acquired for bulk crystals over a range of temperatures. **Figure 5** shows the evolution of bands A-D under increasing temperatures ranging from 4-80 K. With increasing temperature, the relative intensities of band A components (A1 and A2) interplay, where A1 dominates at 4 K, while A2 progressively intensifies at higher temperatures. The increasing temperature drives population from A1 to A2, enhancing A2 and yielding the observed intensity crossover.

An analogous trend is observed in bands C and D, with C1 and D1 predominant at 4 K, while C2 and D2 intensifying with temperature. In band B, an additional component (B2 at 1.797 eV) emerges at 40 K and increases in intensity with temperature. By ~80 K, the individual components of bands A, C and D are no longer discernable, while B1 and B2 exhibit quenching accompanied by spectral broadening. These observations in bands A-D can be explained by enhanced exciton-phonon scattering and the opening of thermally activated non-radiative decay pathways. Notably, these changes occur at temperatures substantially below the Néel temperature ($T_N$ ≈ 120 K), extracted through temperature-dependent susceptibility measurements (SI, **Figure S7**). This clear mismatch indicates that the PL "quenching/decay" temperature does not follow the magnetic ordering transition, suggesting that the observed spectral degradation is dominated by non-radiative and/or phonon-assisted decay channels rather than direct coupling to $T_N$.

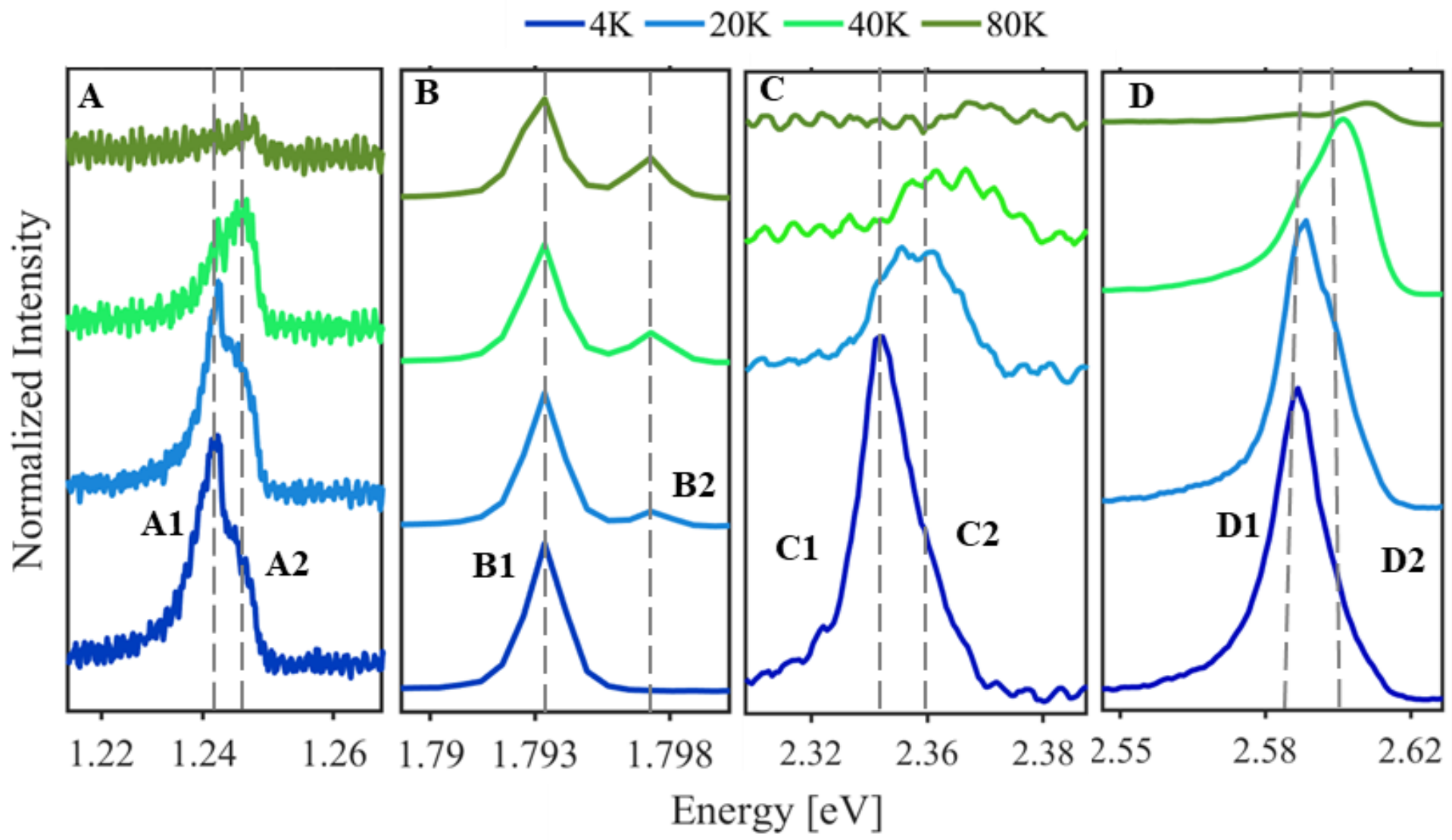

**Figure 5. Emission spectra of bulk $FePS_3$ under increasing temperature.** Evolution of μPL of bands A-D under increasing temperatures ranging from 4-80K.

Moreover, a similar trend in bands B and C was observed in the monolayer limit (see SI, **Figure S10**). This is similar to the polarization responses observed above for these two bands. A further explanation of this phenomenon will be provided in the discussion section. Additionally, the opposing behavior of the components in bands A-D (i.e. peaks A1, A2, C1, C2, D1, D2) under increasing temperature provides further proof of their discrete origins.

## 3. Discussion

To understand the impact of the $FeS_6$ octahedral distortion revealed by single-crystal XRD measurements, as well as the origin of the optical transitions discussed above, density-functional theory (DFT) calculations were performed for bulk, bilayer, and monolayer $FePS_3$. As shown in **Figure 1**, the $FeS_6$ octahedron exhibits three inequivalent Fe–S bond lengths, in contrast to the nearly isotropic $MnS_6$ octahedron in $MnPS_3$, which serves as a reference honeycomb lattice. This distortion leads to inequivalent Fe–Fe nearest-neighbor distances in $FePS_3$ (**Tables S2–S6** in the SI) and reveals a symmetry breaking within the b/a plane. As a consequence, the nearest-neighbor exchange interactions in $FePS_3$ display a significant variance, as detailed in **Table S7** of the SI.

Beyond these structural and magnetic consequences, the distortion directly impacts the electronic structure and optical response of $FePS_3$. The Fe 3d states are primarily split by the octahedral coordination of the surrounding S atoms. However, the distorted $FeS_6$ environment further modifies the orbital character beyond the ideal crystal-field picture. In *the local octahedral reference frame*, where the coordinate axes are aligned with the $FeS_6$ octahedron, the $d_{xz}$ and $d_{yz}$ orbitals remain degenerate, as shown in **Figure S3** in the SI. Conversely, in the *experimental frame,* the Fe–S octahedra are tilted by approximately ~45° with respect to the plane of the $FePS_3$ layers. This tilt induces a strong mixing between different $d$-orbital characters when expressed in $FePS_3$, where the layers lie in the $xy$-plane, leading to a large $d_{z^2}$ contribution. As a result, the effective orbital splitting resembles that expected for a trigonal antiprismatic coordination.[39] The calculated projected density of states confirms this fragmentation of the Fe 3d manifold into several sub-bands, whose relative energy separations remain largely unchanged from bulk to monolayer. These d states define the energetic landscape of the unoccupied states available for the optical transitions described in the results section. While possible intra-atomic d-d excitations (e.g. band A) lie beyond the scope of a single-particle DFT description, bands B-D can be qualitatively associated with families of ligand-to-metal charge-transfer transitions from occupied S 3p states to distinct Fe 3d submanifolds, as shown in the calculated DOS in **Figure 6** (labeled as P1, P2, and P3).

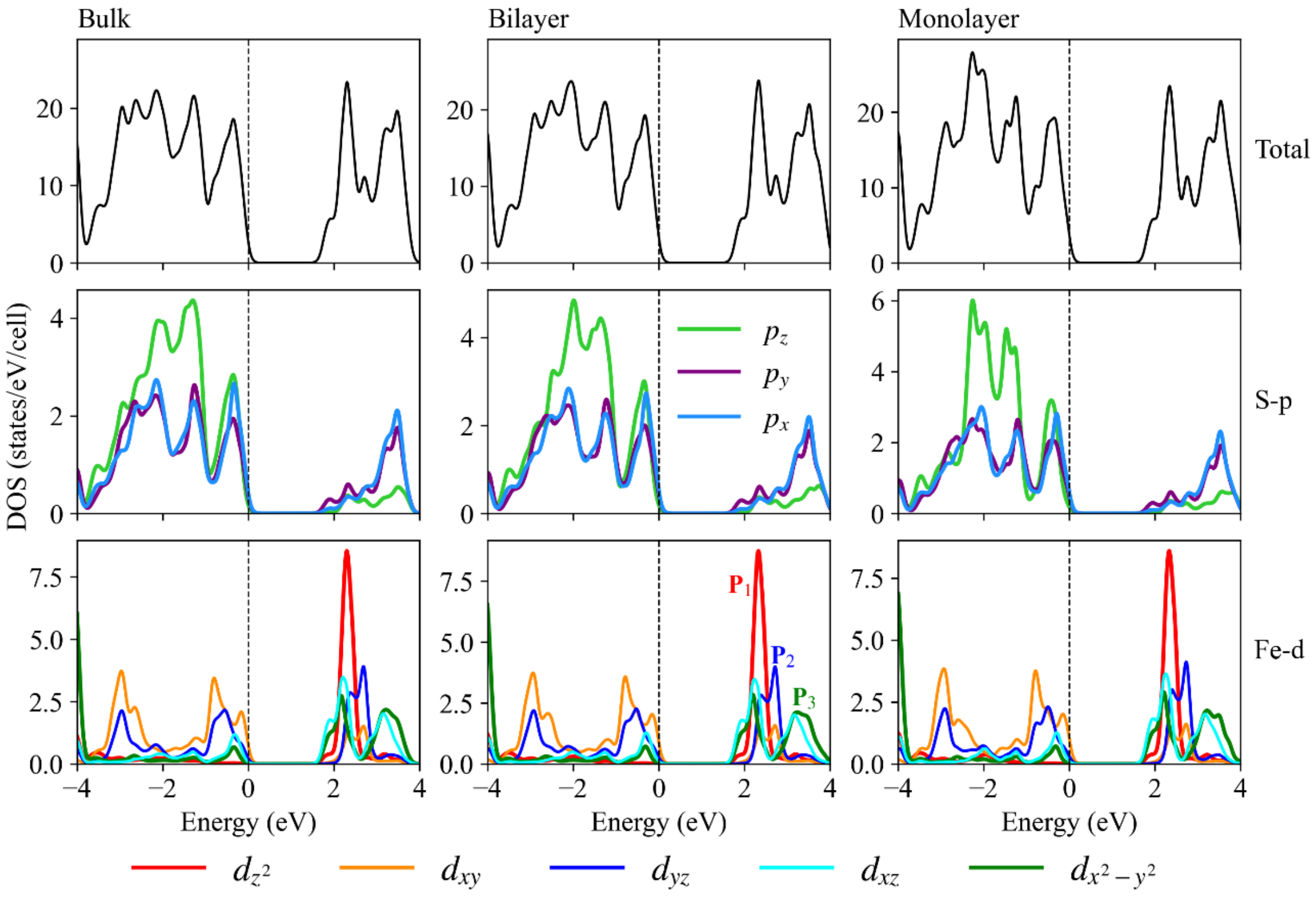

**Figure 6. Total and projected density of states (DOS)**, including the S and Fe atomic contributions resolved into S $3p$ $(p_x, p_y, p_z)$ and Fe $3d$ $(d_{xy}, d_{xz}, d_{yz}, d_{x^2-y^2}, d_{z^2})$ orbitals. The labels P1, P2, and P3 indicate ligand-to-metal charge-transfer transitions from occupied S $3p$ states to distinct Fe $3d$ submanifolds. The total and projected densities of states are shown per unit cell. The DOS corresponds to the sum of spin-up and spin-down contributions.

**Figure 7** reveals the orbital resolved band structure in the $FePS_3$ monolayer case. Here the valence-band edge is primarily composed of S 3p states, with only minor contributions from Fe 3d orbitals, while the conduction bands are dominated by Fe 3d character. Thus, supporting a predominantly *p–d* nature in the dominant optical transitions discussed in the results section. Furthermore, both the valence-band maximum (VBM) and the conduction-band minimum are located at the center of the Brillouin zone (Γ). This facilitates the analysis of dipole-selection rules in terms of momentum matrix elements (MMEs),[40] since the Bloch phase drops out at the Γ point. This is in sharp contrast to transition metal dichalcogenides such as $MoS_2$ which adhere to the selection rules at the K point.[41] The relevance of this distinction is particularly important in relation to the Fe 3d region labeled P1 in **Figure 6** which exhibits a clear $d_{z^2}$ character. This provides an energetically favorable set of final states for optical excitations which can explain band B discussed in the results section. Since the $dz^2$ state is rotational invariant w.r.t. rotations around the z axis and the initial state of the transition is composed of nearly equal amounts of $p_x$ and $p_y$ states, the MMEs for $p_x$ and $p_y$ should be similar, leading to no preferred polarization of the corresponding optical transition. Conversely, the Fe 3d regions labeled P2 and P3 in **Figure 6** exhibit a mixed d character which consists of different contributions from states aligned with the Cartesian x and y axis and, thus, can explain the experimentally observed circular and linear polarization of band C and D. A more detailed analysis of the transitions including the direct calculation of the MMEs is beyond the scope of the present work. It is worth noting, however, that the position of the VBM can shift away from Γ when different van der Waals corrections or Hubbard-U parameters are employed. This pronounced sensitivity to subtle structural details may provide a possible explanation for why certain experimental peaks reported here have not been observed in previous studies.[17,18]

Although the unoccupied Fe 3d submanifolds associated with peaks P1, P2, and P3 remain essentially unchanged from bulk to monolayer (see **Figure 6**), the occupied S 3p states undergo a pronounced reorganization upon exfoliation. This selectively modifies the initial states of the optical transitions, leading to a redistribution and enhanced resolution of the corresponding spectral features in the monolayer, without requiring a substantial rearrangement of the conduction-band electronic structure. However, the rotating of the LP principal axes of the D1 and D2 peaks from the bulk to the monolayer limit may indicate a strong dependence on the reorganization of the S 3p states.

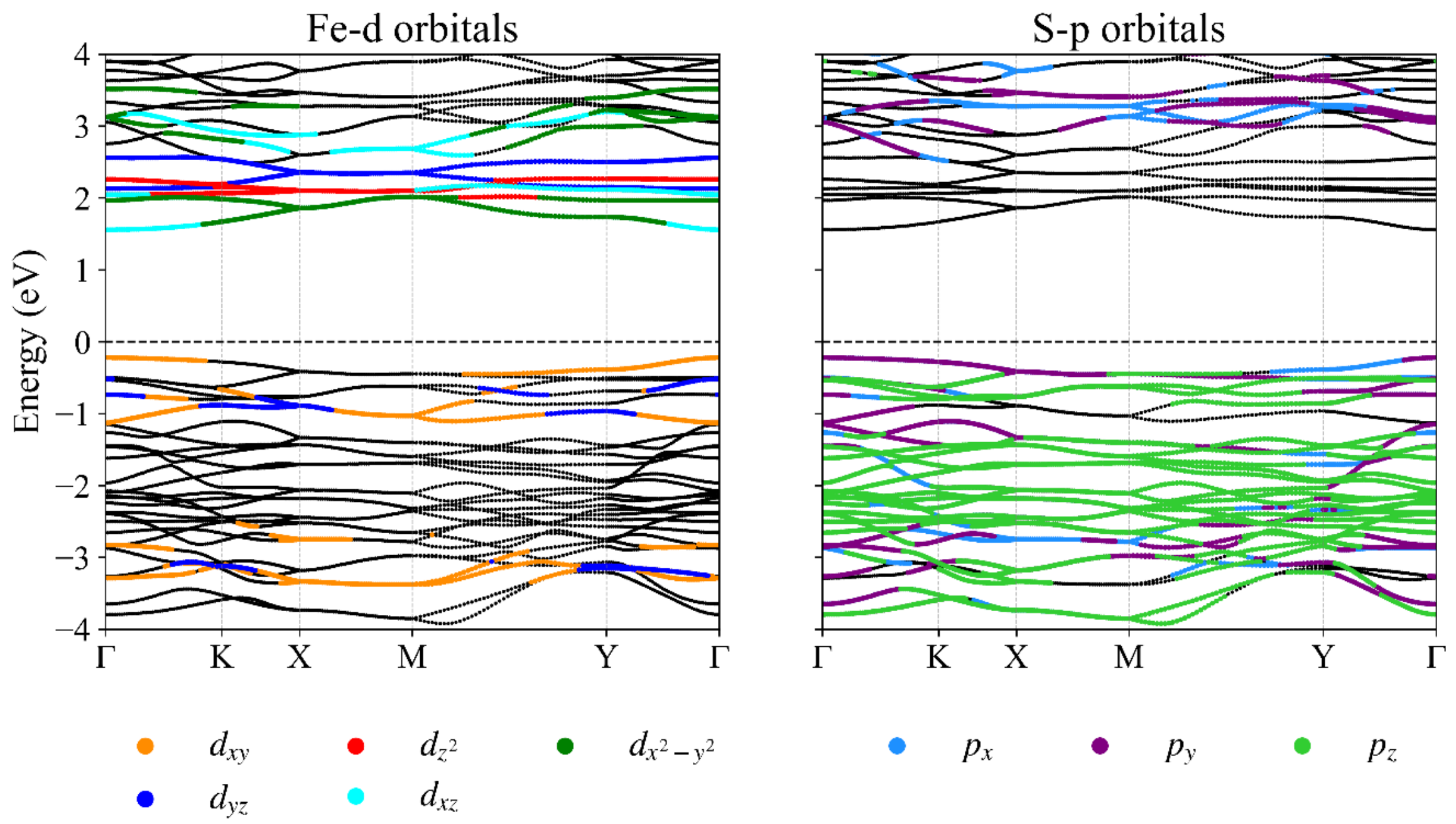

**Figure 7. Orbital-resolved band structure of monolayer $FePS_3$.** The left panel shows the Fe d-orbital contributions, while the right panel shows the S p-orbital contributions. Black points represent the total band structure. Only states with more than 20% d-character and 10% p-character are highlighted, emphasizing the dominant orbital composition of the conduction and valence bands, respectively. A detailed orbital-by-orbital weight analysis is provided in the SI, Figure S4.

## 4. Conclusion

The current study reveals the impact of in-plane structural anisotropy on the optical properties and magnetic ordering of $FePS_3$ from the single-crystal bulk to the monolayer form. Single-crystal XRD measurements uncovered an enlarged a/b lattice-parameter ratio, owing to the previously reported distorted $FeS_6$ octahedron and resulting inequivalent Fe-Fe distance. This, in turn, causes significant changes in its electronic and optical transitions. μPL measurements revealed four emission transitions, one intra-atomic *d-d* transition and three *p-d* charge transfer transitions. The three *p-d* charge transfer transitions revealed different polarization responses, down to the monolayer limit. With the use of DFT calculations, for the first time, an answer for these varied polarization responses was provided. Band B revealed a lack of polarization. This was surprising as previous work on $FePS_3$ reported that the CBM was located at the K point, while the VBM on the K → Γ path.[42] Thus, this result was in contrast to the selection rules at the K point. However, our calculations revealed both the VBM and the CBM to be at the center of the Brillouin zone, i.e. the Γ point. With the clear $dz^2$ character in the electronic structure, labelled P1, the transition is composed of nearly equal amounts of $p_x$ and $p_y$ states. Thus, leading to no preferred polarization of the corresponding optical transition, band B in this case. Conversely, bands C and D revealed a CP and LP, respectively. This can be attributed to the two other Fe 3d regions in our electronic band structure (P2 and P3) which exhibited a mixed d character with different contributions from states aligned with the Cartesian x and y axis. A more detailed analysis of the transitions including the direct calculation of the MMEs is beyond the scope of the present work and will be a topic of ongoing research. Moreover, the main optical responses (i.e., μPL, polarization and temperature dependent μPL) of bands B and C remain unchanged from the bulk to monolayer limit. This is attributed to a reorganization of the S 3p states upon exfoliation without a substantial rearrangement of the CB. However, in band D a rotation of the LP principal axes of the D1 and D2 peaks indicate a strong dependence on the reorganization of the S 3p states. Our work highlights the strong impact structural anisotropy in $FePS_3$ has on its optical responses and magnetic order. We provide additional insight into the origins and contrasting polarization behavior of *p-d* charge transfer transitions in $FePS_3$. Further investigations are required regarding these transitions, and it is an ongoing research endeavor in our group.

## 5. Experimental Section

### Sample preparation

$FePS_3$ bulk crystals were synthesized via a chemical vapor transport (CVT) method described in our previous work.[13] The iron, red phosphorus, and sulfur in the ratio (1:1:3.15) were sealed inside an evacuated quartz ampoule with pressure ~5 x $10^{-5}$ Torr. A 5% molar excess of sulfur was added as a transporting agent to increase the chemical yield. The ampoule was placed in a two-zone furnace where the substrate zone was kept to 850°C and the deposition zone at 790°C for 7 days.

Few-layer and monolayer flakes were prepared via mechanical exfoliation of the single-crystal bulk $FePS_3$ samples. These flakes were encapsulated with mechanically exfoliated hexagonal boron nitride (h-BN; commercially available from 2D Semiconductors) to protect the exfoliated samples from oxidization. Furthermore, the use of h-BN to encapsulate the single-layered material has been reported to significantly improve the PL quantum yield as well as reduce the exciton-exciton annihilation rate, as shown by YongJun Lee *et.al*[1]. The dry-transfer technique was used to cap $FePS_3$ in h-BN, whereby a mechanically exfoliated $FePS_3$ flake consisting of few-layer and monolayer regions was deposited on a $SiO_2$/Si wafer within a glovebox where the oxygen levels can be controlled. A thin h-BN flake larger than the $FePS_3$ flake to allow for complete capping was prepared using mechanical exfoliation. Using an optical microscope, the h-BN flake on a PDMS/PC film was aligned with the desired $FePS_3$ flake on the $SiO_2$/Si wafer and released using heat. Finally, the $SiO_2$/Si wafer containing the h-BN/$FePS_3$ structure was treated with acetone to remove any PDMS/PC film that remained.

**General Characterization of bulk crystals**

*Single-crystal X-ray diffraction measurements*. The crystal structures of bulk $FePS_3$ were determined at room temperature and 140 K using a Rigaku XtaLAB Synergy-S CCD, equipped with an Oxford AD51 Dry air unit series Cryostream system with low temperatures. Measurements were carried out using Mo & Cu irradiation. The single-crystal structures were solved using Olex2.

*SEM and EDX measurements*. SEM images of $FePS_3$ bulk crystals and Cr:$ZnPS_3$ bulk crystals were acquired on a Zeiss Ultra Plus FEG-SEM using an SE2 (Everhart-Thornley) detector at an accelerating voltage of 3.6 kV. EDX spectra were collected with an X-Max silicon drift detector at an accelerating voltage of 12 kV. All SEM-EDX measurements were performed in on bulk crystals mounted on carbon tape, with the working distance maintained at 6-10 mm.

**Optical and Magneto-Optical spectroscopy**

Micro-photoluminescence (μ-PL) spectra were measured by mounting single-crystal bulk $FePS_3$ and h-BN/$FePS_3$ exfoliated structures on a $SiO_2$/Si substrate onto a fiber-based confocal microscope. The microscope included an objective lens with an NA of 0.65 and a 473 nm long-pass dichroic mirror. The microscope was embedded into a cryogenic system (attoDRY1000 closed cycle cryostat) to allow for optical measurements at 4K. A ceramic heater on the sample stage allowed for temperature-dependent μ-PL spectra to be recorded, where the temperature of the sample increased gradually from 4K-200K. A magnet within the cryostat system enabled magneto-μPL (MPL) spectra to be recorded where the external magnetic field varied from 0T-7T and -7T. Circular polarized μ-PL and MPL and spectra were recorded using a suitable linear polarizer and λ/4 waveplate in line with the emitted light from the sample. Linear polarized μ-PL and MPL spectra were recorded by placing a linear polarizer in the emission line of the confocal microscope, which were rotated from 0°-360°. The target sample was excited using a continuous wave 405 nm laser diode, where the emission from the sample was detected using a FERGIE spectrograph. XYZ piezo-electric positioners allowed to focus the laser beam on a desired spot on the sample. All optical measurements were recorded under an illumination power of 0.3 mW.

**Computational details**

<u>Exchange couplings:</u> The calculations were carried out using the density functional theory (DFT) applying the Perdew–Burke–Ernzerhof (PBE) generalized gradient approximation to express the exchange-correlation functional[43]using the projector augmented wave (PAW) pseudopotentials, as implemented in the open-source DFT package Quantum Espresso.[44,45] All calculations were carried out with a plane-wave basis using a kinetic energy cut-off of 80 and 90 Ry for $FePS_3$ and $MnPS_3$, respectively. To account for the d-orbital localization, we incorporate the Hubbard potential (DFT+U), setting U as 3 eV for 3d open shells of Mn and Fe atoms.[16] To optimize the structure of $MPX_3$ monolayers, we utilized a Birch-Murnaghan plot while maintaining a constant a/b lattice parameters ratio, as determined through single crystal XRD measurements. Following the cell optimization process, ion relaxation was employed to ensure complete structural optimization of the monolayers. The exchange interaction variables of anisotropic $FePS_3$ were extracted using DFT+U. To extract the J-coefficients, all calculations were spin-polarized and were set to 7 magnetic supercell configurations, all of which are shown in **Figure 8**. Using the DFT energies of these magnetic configurations, we were able to extract a set of six linear, independent equations to calculate all of the anisotropic exchange parameters in $FePS_3$ presented in **Equation 2**:

$$
\begin{gathered}
J_{11} = \frac{E_{FM} - E_{N\acute{e}el} - E_{stripe} + E_{zigzag}}{8S^2} \\
J_{21} = \frac{E_{FM} + E_{N\acute{e}el} - E_{stripe} - E_{zigzag}}{16S^2} \\
J_{31} = \frac{E_{FM} - E_{N\acute{e}el} + E_{stripe} - E_{zigzag} + 2\left(E_{abzigzag} - 2E_{stripezigzag} + E_{abstripe}\right)}{16S^2} \\
J_{12} = \frac{E_{FM} - E_{N\acute{e}el} + E_{stripe} - E_{zigzag} + 2\left(E_{abzigzag} - E_{abstripe}\right)}{8S^2} \\
J_{22} = \frac{E_{FM} + E_{N\acute{e}el} + E_{stripe} + E_{zigzag} - 2\left(E_{abzigzag} + E_{abstripe}\right)}{16S^2} \\
J_{32} = \frac{E_{stripezigzag} - E_{abzigzag}}{2S^2}
\end{gathered}
\tag{2}
$$

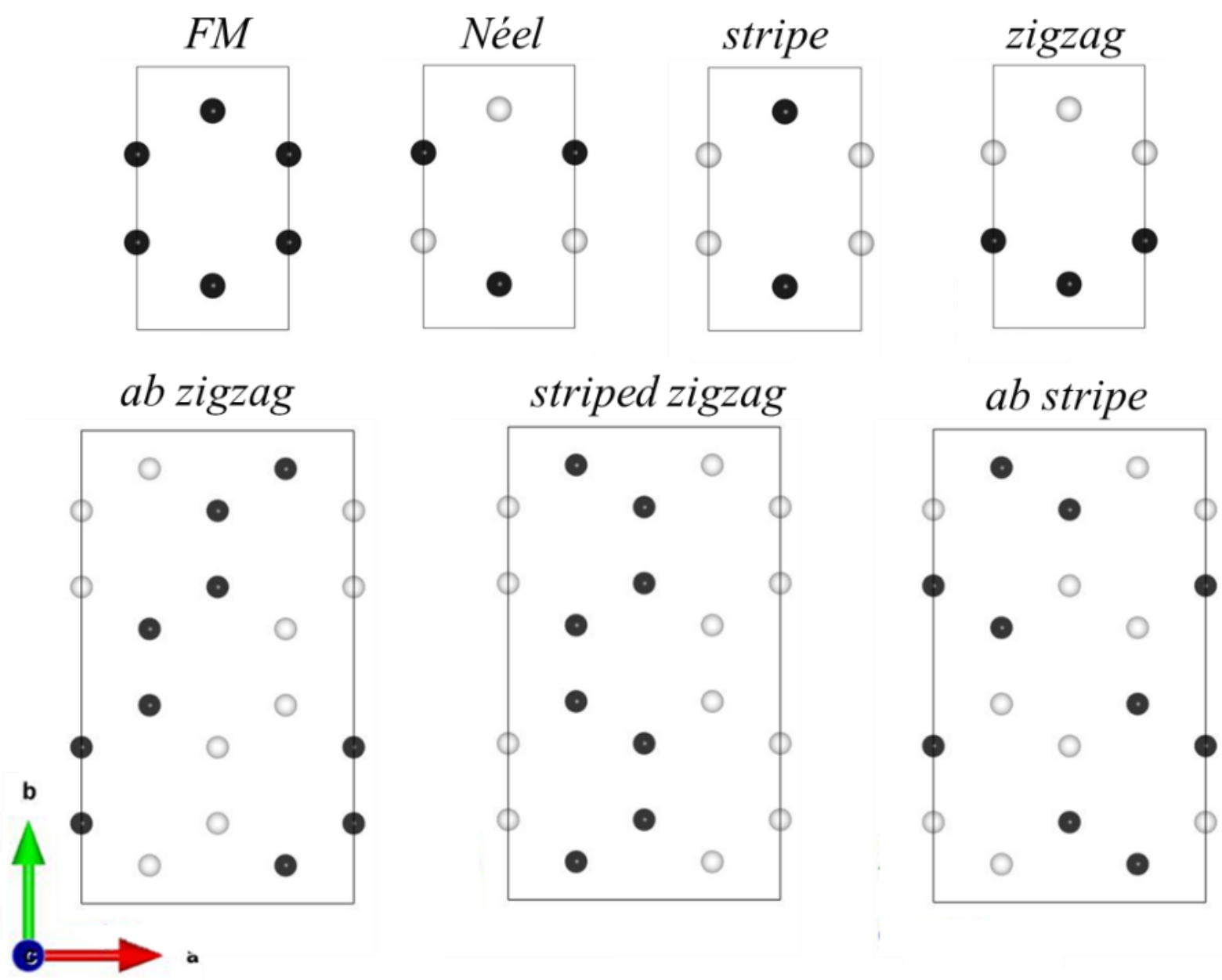

**Figure 8. Seven magnetic supercell configurations.** Black and white atoms refer to Fe in either a spin up or down state.

To validate our results, we tested our calculations against an isotropic $FePS_3$ structure. For this case, the correct Heisenberg Hamiltonian collapses into tri-degenerate exchange interactions as described by **Equation 3**:

$$J_1 = \frac{E_{FM} - E_{N\acute{e}el} - E_{stripe} + E_{zigzag}}{8S^2}$$
$$J_2 = \frac{E_{FM} + E_{N\acute{e}el} - E_{stripe} - E_{zigzag}}{16S^2} \quad (3)$$
$$J_3 = \frac{E_{FM} - E_{N\acute{e}el} + 3E_{stripe} - 3E_{zigzag}}{24S^2}$$

Inversion symmetry breaking: We quantify the degree of inversion symmetry breaking by the $\Delta_{a,b,c}$ parameter, as presented by **Equation 4**:

$$\Delta_a = \left(\overline{MS_{1,2}} - \overline{MS_{4,5}}\right)/\overline{MS}$$
$$\Delta_b = \left(\overline{MS_{1,5}} - \overline{MS_{2,4}}\right)/\overline{MS} \quad (4)$$
$$\Delta_c = (MS_6 - MS_3)/\overline{MS}$$

A perfect octahedron with equal MS bond lengths will compute to be $\Delta_{a,b,c} = 0$ where any deviation from it results in inversion symmetry breaking. The $\Delta_{a,b,c}$ parameters are a projected quantity over the a, b, and c crystallographic axis, respectively. Therefore, the inversion offset can be defined as the magnitude of $\|\Delta_a, \Delta_b, \Delta_c\|$. However, we find that $\Delta_a$ and $\Delta_c$ are negligible compared to $\Delta_b$, hence we focus only on the $\Delta_b$ parameter.

Electronic band structure: The calculations were performed within DFT using the PBE generalized gradient approximation for the exchange–correlation functional,[46] as implemented in the Vienna *Ab initio* Simulation Package (VASP).[47] The interaction between valence electrons and ionic cores was described using the projector augmented-wave (PAW) method. Structural relaxations were carried out until the residual forces on each atom were smaller than $10^{-3}$ Å$^{-1}$, using a plane-wave energy cutoff of 500 eV. Regular Monkhorst–Pack $k$-point meshes of $11 \times 5 \times 1$ for structural relaxations and $15 \times 9 \times 1$ for self-consistent calculations were employed for the monolayer and bilayer systems, while meshes of $11 \times 5 \times 3$ and $15 \times 9 \times 5$ were used for the bulk case. The calculations were performed using the experimental bulk lattice parameters $a$ = 5.947 Å, $b$ = 10.30 Å, and $c$ = 6.72 Å, as reported in Ref.[13] Atomic positions were subsequently relaxed. The same relaxation procedure was adopted for the reduced-dimensionality systems.

To account for the localized nature of the Fe $3d$ orbitals, on-site Coulomb interactions were included within the DFT+$U$ formalism, with $U = 3.2$ eV applied to the Fe atoms. Long-range van der Waals interactions were treated using the Grimme D3 dispersion correction,[48] which provides an accurate description of interatomic distances and yields structural parameters in good agreement with experimental data. The electronic density of states was calculated using a Gaussian smearing scheme.

**Acknowledgements**

The authors gratefully acknowledge the computing time made available to them on the high-performance computer at the NHR Center of TU Dresden. This center is jointly supported by the Federal Ministry of Research, Technology and Space of Germany and the state governments participating in the NHR (www.nhr-verein.de/unsere-partner). A.K.B. and E.L. were supported by the European Commission via the Marie-Sklodowska Curie action Phonsi (H2020-MSCA-ITN-642656). E.L. thanks the support of Deutsch-Israelische Projektkooperation (DIP, Project No. NA1223/2-1) and the Israel Science Foundation (ISF, Project No. 149/23). The authors are grateful to Prof. Magdalena Birowska and her student Mr. Milosz Marcin Rybak for their initial work on the theoretical analysis of this work which paved the way to the authors' theoretical findings. The authors thank Dr. Natalia Fridman for her assistance with single-crystal XRD measurements. The authors are also grateful to Ms. Adi Harchol for assistance with the synthesis of Cr:$ZnPS_3$ and to Ms. Diksha Prabhu Gaonkar for performing EDX measurements on them.

**Conflict of Interest**

Authors have no conflicts of interest to declare.

**References**

(1) Chittari, B. L.; Park, Y.; Lee, D.; Han, M.; MacDonald, A. H.; Hwang, E.; Jung, J. Electronic and Magnetic Properties of Single-Layer $MPX_3$ Metal Phosphorous Trichalcogenides. *Phys. Rev. B* **2016**, *94*, 184428.

(2) Yang, J.; Zhou, Y.; Guo, Q.; Dedkov, Y.; Voloshina, E. Electronic, Magnetic and Optical Properties of $MnPX_3$ (X = S, Se) Monolayers with and without Chalcogen Defects: A First-Principles Study. *RSC Adv.* **2020**, *10*, 851–864.

(3) Li, X.; Cao, T.; Niu, Q.; Shi, J.; Feng, J. Coupling the Valley Degree of Freedom to Antiferromagnetic Order. *Proceedings of the National Academy of Sciences* **2013**, *110*, 3738–3742.

(4) Pei, Q.; Wang, X.; Zou, J.; Mi, W. Half-Metallicity and Spin-Valley Coupling in 5d Transition Metal Substituted Monolayer $MnPSe_3$. *J. Mater. Chem. C Mater.* **2018**, *6*, 8092–8098.

(5) Wildes, A. R.; Simonet, V.; Ressouche, E.; McIntyre, G. J.; Avdeev, M.; Suard, E.; Kimber, S. A. J.; Lançon, D.; Pepe, G.; Moubaraki, B.; Hicks, T. J. Magnetic Structure of the Quasi-Two-Dimensional Antiferromagnet NiPS3. *Phys. Rev. B Condens. Matter Mater. Phys.* **2015**, *92*.

(6) Brec, R. Review on Structural and Chemical Properties of Transition Metal Phosphorous Trisulfides $MPS_3$. *Solid State Ion.* **1986**, *22*, 3–30.

(7) Kurosawa, K.; S. S.; Y. Y. Neutron Diffraction Study on $MnPS_3$ and $FePS_3$. *J. Physical Soc. Japan* **1983**, 3919–1926.

(8) Joy, P. A.; Vasudevan, S. Magnetism in the Layered Transition-Metal Thiophosphates MPS3 *(M =Mn, Fe, and Ni)*; *Phys. Rev. B 46, 5425*, ***1992****; Vol. 46.*

(9) Joy, P. A.; Vasudevan, S. Optical-Absorption Spectra of the Layered Transition-Metal Thiophosphates $MPS_3$ (M=Mn, Fe, and Ni); *Phys. Rev. B 46, 5134,* **1992**; *Vol. 8.*

(10) Chittari, B. L.; Park, Y.; Lee, D.; Han, M.; Macdonald, A. H.; Hwang, E.; Jung, J. Electronic and Magnetic Properties of Single-Layer $MPX_3$ Metal Phosphorous Trichalcogenides. *Phys. Rev. B* **2016**, *94*.

(11) Sivadas, N.; Daniels, M. W.; Swendsen, R. H.; Okamoto, S.; Xiao, D. Magnetic Ground State of Semiconducting Transition-Metal Trichalcogenide Monolayers. *Phys. Rev. B Condens. Matter Mater. Phys.* **2015**, *91*.

(12) Mayorga-Martinez, C. C.; Sofer, Z.; Sedmidubský, D.; Huber, Š.; Eng, A. Y. S.; Pumera, M. Layered Metal Thiophosphite Materials: Magnetic, Electrochemical, and Electronic Properties. *ACS Appl. Mater. Interfaces* **2017**, *9*, 12563–12573.

(13) Budniak, A. K.; Zelewski, S. J.; Birowska, M.; Woźniak, T.; Bendikov, T.; Kauffmann, Y.; Amouyal, Y.; Kudrawiec, R.; Lifshitz, E. Spectroscopy and Structural Investigation of Iron Phosphorus Trisulfide—$FePS_3$. *Adv. Opt. Mater.* **2022**, *10*.

(14) Murayama, C.; Okabe, M.; Urushihara, D.; Asaka, T.; Fukuda, K.; Isobe, M.; Yamamoto, K.; Matsushita, Y. Crystallographic Features Related to a van Der Waals Coupling in the Layered Chalcogenide FePS3. *J. Appl. Phys.* **2016**, *120*.
(15) Zhang, J.; Nie, Y.; Wang, X.; Xia, Q.; Guo, G. Strain Modulation of Magnetic Properties of Monolayer and Bilayer FePS3 Antiferromagnet. *J. Magn. Magn. Mater.* **2021**, *525*, 167687.
(16) Olsen, T. Magnetic Anisotropy and Exchange Interactions of Two-Dimensional FePS3, NiPS3 and MnPS3 from First Principles Calculations. *J. Phys. D Appl. Phys.* **2021**, *54*, 314001.
(17) Lee, Y.; Son, S.; Kim, C.; Kang, S.; Shen, J.; Kenzelmann, M.; Delley, B.; Savchenko, T.; Parchenko, S.; Na, W.; Choi, K.; Kim, W.; Cheong, H.; Derlet, P. M.; Kleibert, A.; Park, J. Giant Magnetic Anisotropy in the Atomically Thin van Der Waals Antiferromagnet $FePS_3$. *Adv. Electron. Mater.* **2023**, *9*.
(18) Zhang, Q.; Hwangbo, K.; Wang, C.; Jiang, Q.; Chu, J.-H.; Wen, H.; Xiao, D.; Xu, X. Observation of Giant Optical Linear Dichroism in a Zigzag Antiferromagnet $FePS_3$. *Nano Lett.* **2021**, *21*, 6938–6945.
(19) Koo, H.-J.; Kremer, R.; Whangbo, M.-H. Unusual Spin Exchanges Mediated by the Molecular Anion P2S64−: Theoretical Analyses of the Magnetic Ground States, Magnetic Anisotropy and Spin Exchanges of $MPS_3$ (M = Mn, Fe, Co, Ni). *Molecules* **2021**, *26*, 1410.
(20) Nitschke, J. E.; Sternemann, L.; Gutnikov, M.; Schiller, K.; Coronado, E.; Omar, A.; Zamborlini, G.; Saraceno, C.; Stupar, M.; Ruiz, A. M.; Esteras, D. L.; Baldoví, J. J.; Anders, F.; Cinchetti, M. Tracing the Ultrafast Buildup and Decay of D-d Transitions in FePS3. *Newton* **2025**, *1*, 100019.
(21) Pestka, B.; Strasdas, J.; Bihlmayer, G.; Budniak, A. K.; Liebmann, M.; Leuth, N.; Boban, H.; Feyer, V.; Cojocariu, I.; Baranowski, D.; Mearini, S.; Amouyal, Y.; Waldecker, L.; Beschoten, B.; Stampfer, C.; Plucinski, L.; Lifshitz, E.; Kratzer, P.; Morgenstern, M. Identifying Band Structure Changes of FePS3 across the Antiferromagnetic Phase Transition. **2024**.
(22) Ghosh, A.; Birowska, M.; Ghose, P. K.; Rybak, M.; Maity, S.; Ghosh, S.; Das, B.; Dey, K.; Bera, S.; Bhardwaj, S.; Nandi, S.; Datta, S. Anisotropic Magnetodielectric Coupling in Layered Antiferromagnetic $FePS_3$. *Phys. Rev. B* **2023**, *108*, L060403.
(23) Chen, J.; Xie, X.; Oyang, X.; Li, S.; He, J.; Liu, Z.; Wang, J.; Liu, Y. Giant Optical Anisotropy Induced by Magnetic Order in $FePS_3$ /$WSe_2$ Heterostructures. *Small* **2024**, *20*.
(24) Ramos, M.; Marques-Moros, F.; Esteras, D. L.; Mañas-Valero, S.; Henríquez-Guerra, E.; Gadea, M.; Baldoví, J. J.; Canet-Ferrer, J.; Coronado, E.; Calvo, M. R. Photoluminescence Enhancement by Band Alignment Engineering in $MoS_2$ /$FePS_3$ van Der Waals Heterostructures. *ACS Appl. Mater. Interfaces* **2022**, *14*, 33482–33490.
(25) Mertens, F.; Mönkebüscher, D.; Parlak, U.; Boix-Constant, C.; Mañas-Valero, S.; Matzer, M.; Adhikari, R.; Bonanni, A.; Coronado, E.; Kalashnikova, A. M.; Bossini, D.; Cinchetti, M. Ultrafast Coherent THz Lattice Dynamics Coupled to Spins in the van Der Waals Antiferromagnet $FePS_3$. *Advanced Materials* **2023**, *35*.

(26) Hwangbo, K.; Zhang, Q.; Jiang, Q.; Wang, Y.; Fonseca, J.; Wang, C.; Diederich, G. M.; Gamelin, D. R.; Xiao, D.; Chu, J.-H.; Yao, W.; Xu, X. Highly Anisotropic Excitons and Multiple Phonon Bound States in a van Der Waals Antiferromagnetic Insulator. *Nat. Nanotechnol.* **2021**, *16*, 655–660.

(27) Kang, S.; Kim, K.; Kim, B. H.; Kim, J.; Sim, K. I.; Lee, J.-U.; Lee, S.; Park, K.; Yun, S.; Kim, T.; Nag, A.; Walters, A.; Garcia-Fernandez, M.; Li, J.; Chapon, L.; Zhou, K.-J.; Son, Y.-W.; Kim, J. H.; Cheong, H.; Park, J.-G. Coherent Many-Body Exciton in van Der Waals Antiferromagnet NiPS3. *Nature* **2020**, *583*, 785–789.

(28) Aruchamy, A.; Berger, H.; Levy, F. Photoelectronic Properties of the P-Type Layered Trichalcogenophosphates FePS3 and FePSe3. *J. Solid State Chem.* **1988**, *72*, 316–323.

(29) Joy, P. A.; Vasudevan, S. Magnetism in the Layered Transition-Metal Thiophosphates MPS3 (M=Mn, Fe, and Ni). *Phys. Rev. B* **1992**, *46*, 5425–5433.

(30) Harms, N. C.; Kim, H.-S.; Clune, A. J.; Smith, K. A.; O'Neal, K. R.; Haglund, A. V.; Mandrus, D. G.; Liu, Z.; Haule, K.; Vanderbilt, D.; Musfeldt, J. L. Piezochromism in the Magnetic Chalcogenide MnPS3. *NPJ Quantum Mater.* **2020**, *5*, 56.

(31) Xu, J.; Liu, C.; Huang, C.; Zheng, H.; Chen, G.; Fan, J.; Zhu, Y.; Ma, C. Influence of Strain on the Magnetic and Optical Properties of Monolayer $XPS_3$. *Opt. Express* **2025**, *33*, 41511.

(32) Cao, X.; Zhang, D.; Ye, W.; Zhou, J.; Zheng, C.; Watanabe, K.; Taniguchi, T.; Ning, J.; Xu, S. Dielectric Screening Effects in the Decoherence of Excitons and Exciton-Phonon Scattering in Atomical Monolayer $WS_2$ Semiconductors. *Phys. Rev. B* **2025**, *112*, 045420.

(33) S. Latini; T. Olsen; K. S. Thygesen. Excitons in van Der Waals Heterostructures: The Important Role of Dielectric Screening. *Phys. Rev. B* **2015**, *92*, 245123.

(34) Li, L. H.; Santos, E. J. G.; Xing, T.; Cappelluti, E.; Roldán, R.; Chen, Y.; Watanabe, K.; Taniguchi, T. Dielectric Screening in Atomically Thin Boron Nitride Nanosheets. *Nano Lett.* **2015**, *15*, 218–223.

(35) Lone, R. H.; Gaonkar, S.; Kumar, B. M.; Kannan, E. S. Manipulation of Trions to Enhance the Excitonic Emission in Monolayer P-$MoS_2$ and Its Hetero-Bilayer by Reverse Charge Injection. *Nanoscale* **2025**, *17*, 1473–1483.

(36) Zhao, H.; Zhao, Y.; Song, Y.; Zhou, M.; Lv, W.; Tao, L.; Feng, Y.; Song, B.; Ma, Y.; Zhang, J.; Xiao, J.; Wang, Y.; Lien, D.-H.; Amani, M.; Kim, H.; Chen, X.; Wu, Z.; Ni, Z.; Wang, P.; Shi, Y.; Ma, H.; Zhang, X.; Xu, J.-B.; Troisi, A.; Javey, A.; Wang, X. Strong Optical Response and Light Emission from a Monolayer Molecular Crystal. *Nat. Commun.* **2019**, *10*, 5589.

(37) Lee, Y.; Forte, J. D. S.; Chaves, A.; Kumar, A.; Tran, T. T.; Kim, Y.; Roy, S.; Taniguchi, T.; Watanabe, K.; Chernikov, A.; Jang, J. I.; Low, T.; Kim, J. Boosting Quantum Yields in Two-Dimensional Semiconductors via Proximal Metal Plates. *Nat. Commun.* **2021**, *12*, 7095.

(38) Tan, Q.; Luo, W.; Li, T.; Cao, J.; Kitadai, H.; Wang, X.; Ling, X. Charge-Transfer-Enhanced $d-d$ Emission in Antiferromagnetic NiPS3. *Appl. Phys. Rev.* **2022**, *9*.

(39) Autieri, C.; Cuono, G.; Noce, C.; Rybak, M.; Kotur, K. M.; Agrapidis, C. E.; Wohlfeld, K.; Birowska, M. Limited Ferromagnetic Interactions in Monolayers of $MPS_3$ (M = Mn and Ni). *The Journal of Physical Chemistry C* **2022**, *126*, 6791–6802.

(40) Woźniak, T.; Faria Junior, P. E.; Seifert, G.; Chaves, A.; Kunstmann, J. Exciton Factors of van Der Waals Heterostructures from First-Principles Calculations. *Phys. Rev. B* **2020**, *101*, 235408.
(41) Hong, J.; Hu, Z.; Probert, M.; Li, K.; Lv, D.; Yang, X.; Gu, L.; Mao, N.; Feng, Q.; Xie, L.; Zhang, J.; Wu, D.; Zhang, Z.; Jin, C.; Ji, W.; Zhang, X.; Yuan, J.; Zhang, Z. Exploring Atomic Defects in Molybdenum Disulphide Monolayers. *Nat. Commun.* **2015**, *6*, 6293.
(42) Xu, D.; Guo, Z.; Tu, Y.; Li, X.; Chen, Y.; Chen, Z.; Tian, B.; Chen, S.; Shi, Y.; Li, Y.; Su, C.; Fan, D. Controllable Nonlinear Optical Properties of Different-Sized Iron Phosphorus Trichalcogenide ($FePS_3$ ) Nanosheets. *Nanophotonics* **2020**, *9*, 4555–4564.
(43) Perdew, J. P.; Burke, K.; Ernzerhof, M. Generalized Gradient Approximation Made Simple. *Phys. Rev. Lett.* **1996**, *77*, 3865–3868.
(44) Giannozzi, P.; Baroni, S.; Bonini, N.; Calandra, M.; Car, R.; Cavazzoni, C.; Ceresoli, D.; Chiarotti, G. L.; Cococcioni, M.; Dabo, I.; Dal Corso, A.; de Gironcoli, S.; Fabris, S.; Fratesi, G.; Gebauer, R.; Gerstmann, U.; Gougoussis, C.; Kokalj, A.; Lazzeri, M.; Martin-Samos, L.; Marzari, N.; Mauri, F.; Mazzarello, R.; Paolini, S.; Pasquarello, A.; Paulatto, L.; Sbraccia, C.; Scandolo, S.; Sclauzero, G.; Seitsonen, A. P.; Smogunov, A.; Umari, P.; Wentzcovitch, R. M. QUANTUM ESPRESSO: A Modular and Open-Source Software Project for Quantum Simulations of Materials. *Journal of Physics: Condensed Matter* **2009**, *21*, 395502.
(45) Giannozzi, P.; Andreussi, O.; Brumme, T.; Bunau, O.; Buongiorno Nardelli, M.; Calandra, M.; Car, R.; Cavazzoni, C.; Ceresoli, D.; Cococcioni, M.; Colonna, N.; Carnimeo, I.; Dal Corso, A.; de Gironcoli, S.; Delugas, P.; DiStasio, R. A.; Ferretti, A.; Floris, A.; Fratesi, G.; Fugallo, G.; Gebauer, R.; Gerstmann, U.; Giustino, F.; Gorni, T.; Jia, J.; Kawamura, M.; Ko, H.-Y.; Kokalj, A.; Küçükbenli, E.; Lazzeri, M.; Marsili, M.; Marzari, N.; Mauri, F.; Nguyen, N. L.; Nguyen, H.-V.; Otero-de-la-Roza, A.; Paulatto, L.; Poncé, S.; Rocca, D.; Sabatini, R.; Santra, B.; Schlipf, M.; Seitsonen, A. P.; Smogunov, A.; Timrov, I.; Thonhauser, T.; Umari, P.; Vast, N.; Wu, X.; Baroni, S. Advanced Capabilities for Materials Modelling with Quantum ESPRESSO. *Journal of Physics: Condensed Matter* **2017**, *29*, 465901.
(46) Perdew, J. P.; Burke, K.; Ernzerhof, M. Generalized Gradient Approximation Made Simple. *Phys. Rev. Lett.* **1996**, *77*, 3865–3868.
(47) Kresse, G.; Joubert, D. From Ultrasoft Pseudopotentials to the Projector Augmented-Wave Method. *Phys. Rev. B* **1999**, *59*, 1758–1775.
(48) Grimme, S.; Antony, J.; Ehrlich, S.; Krieg, H. A Consistent and Accurate *Ab Initio* Parametrization of Density Functional Dispersion Correction (DFT-D) for the 94 Elements H-Pu. *J. Chem. Phys.* **2010**, *132*.

## Supporting Information

**Crystal anisotropy implications on the intrinsic magneto-optical properties of van der Waals $FePS_3$**

*Ellenor Geraffy[1#], Kusha Sharma,[1#] Shahar Zuri[1], Faris Horani[1,2], Adam K. Budniak[1], Muhamed Dawod[3], Yaron Amouyal[3], Thomas Brumme[4] Andrea Maricel León[7], Thomas Heine[4,5,6*], Rajesh Kumar[8], Doron Naveh[8], and Efrat Lifshitz[1*]*

[1]Schulich Faculty of Chemistry, Solid State Institute, Russell Berrie Nanotechnology Institute, and the Helen Diller Quantum Information Center, Technion – Israel Institute of Technology, 3200003 Haifa, Israel

[2]Department of Chemistry, University of Washington, Seattle, Washington 98195-1700, United States

[3]Department of Materials Science and Engineering, Technion – Israel Institute of Technology, 3200003 Haifa, Israel

[4]Faculty of Chemistry and Food Chemistry, Technical University of Dresden, 01069, Dresden, Germany

[5]Center for Advanced Systems Understanding, CASUS, HZDR, 02826 Görlitz, Germany

[6]Department of Chemistry and IBS for nanomedicine, Yonsei University, Seoul 120-749, Republic of Korea

[7]Departamento de Física, Facultad de Ciencias, Universidad de Chile, Casilla 653, Santiago, Chile

[8]Electrical Engineering, Bar Ilan University, Ramat Gan, Israel

#Equal contribution
*Corresponding authors
Email: ssefrat@technion.ac.il, thomas.heine@tu-dresden.de

# 1. Characterization of single crystal $FePS_3$

## 1.1. *Energy-dispersive X-ray spectroscopy measurements of bulk (parent) crystal*

To ascertain stoichiometric composition and purity of the fabricated $FePS_3$ bulk crystals, energy-dispersive X-ray spectroscopy (EDX) measurements were carried out on all samples discussed in the main text. Figure S1 displays the full EDX spectrum of the $FePS_3$ bulk crystal, with the inset showing a scanning electron microscope (SEM) image of one of the samples. Table S1 shows the composition of the bulk crystal, which shows the expected 1:1:3 (Fe:P:S) ratio with no contaminants.

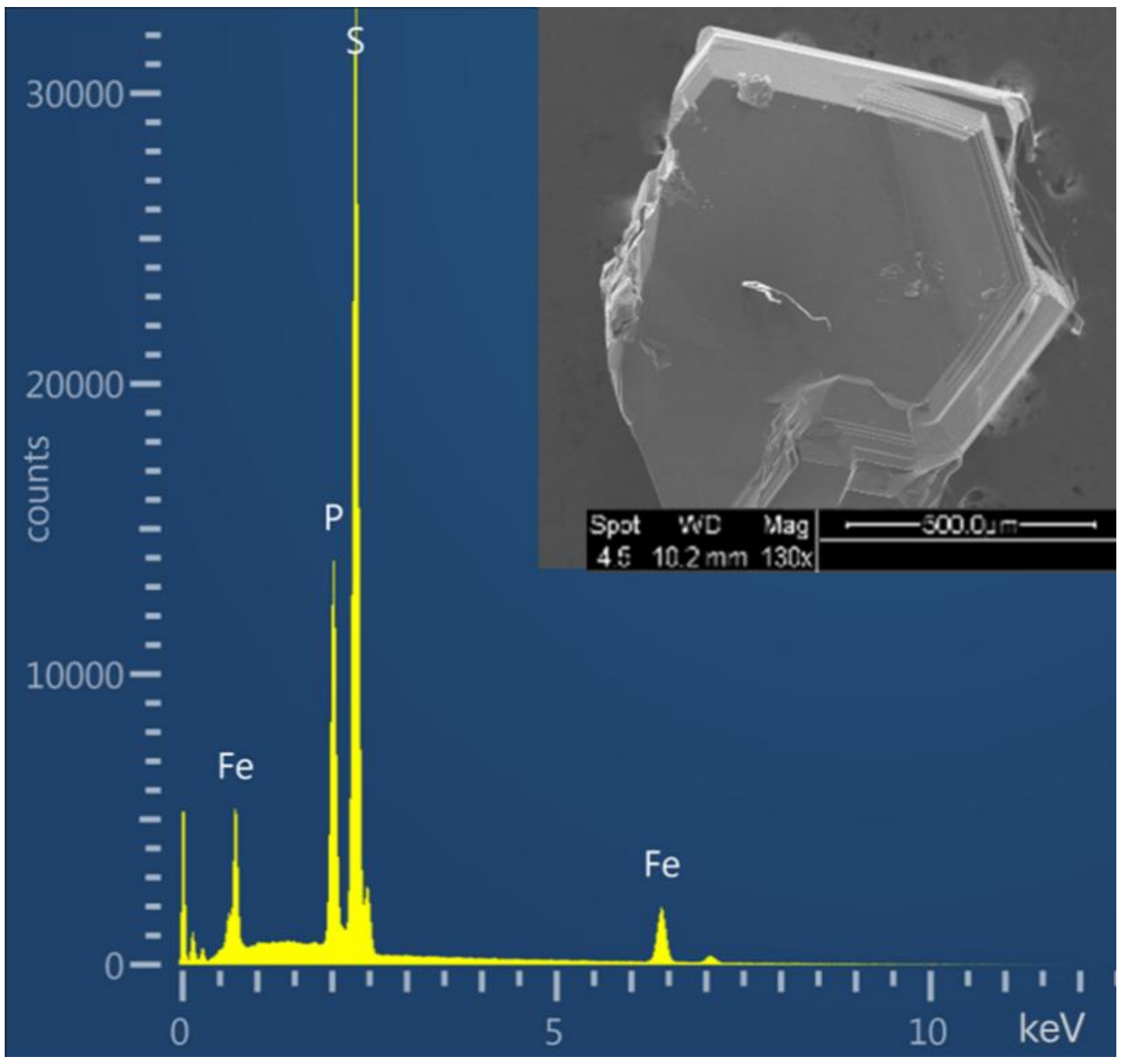

**Figure S1. Characterization of bulk FePS3 single crystal.** EDX spectrum, showing the elements that make up the crystal. Inset: SEM image of the crystal.

**Table S1: Composition of bulk $FePS_3$ crystal from EDX measurements**

| Element | Atomic% | At. % sigma (+/-) |
|---|---|---|
| P | 20.49 | 0.17 |
| S | 60.15 | 0.23 |
| Fe | 19.36 | 0.22 |
| Total | 100.00 | |

## 1.2. *XRD measurements of the bulk (parent) crystal*

The crystallinity of the bulk crystal at a lower temperature was verified using single-crystal X-ray diffraction (XRD) measurement. A full set of XRD data is given in Table S2-5. These results provided benchmark parameters for all DFT calculations which will be elaborated on in the next section.

**Table S2: Bond lengths in bulk $FePS_3$ at 140K**

| Atom | Atom | Length/Å | Atom | Atom | Length/Å |
|---|---|---|---|---|---|
| Fe1 | S1 | 2.5361(16) | S1 | P1 | 2.0244(17) |
| Fe1 | $S1^{1}$ | 2.5442(14) | S2 | $Fe1^{4}$ | 2.5464(17) |
| Fe1 | $S1^{2}$ | 2.5442(14) | S2 | $Fe1^{6}$ | 2.5464(17) |
| Fe1 | $S1^{3}$ | 2.5361(16) | S2 | P1 | 2.031(3) |
| Fe1 | $S2^{4}$ | 2.5464(17) | P1 | $S1^{7}$ | 2.0244(17) |
| Fe1 | $S2^{5}$ | 2.5464(17) | P1 | $P1^{8}$ | 2.182(4) |
| S1 | $Fe1^{1}$ | 2.5442(14) | | | |

**Table S3: Bond angles in bulk $FePS_3$ at 140K**

| Atom | Atom | Atom | Angle/° | Atom | Atom | Atom | Angle/° |
|---|---|---|---|---|---|---|---|
| S1 | Fe1 | $S1^{1}$ | 94.91(7) | $S2^{4}$ | Fe1 | $S2^{5}$ | 94.85(7) |
| $S1^{1}$ | Fe1 | $S1^{2}$ | 95.51(5) | Fe1 | S1 | $Fe1^{3}$ | 84.49(5) |
| S1 | Fe1 | $S1^{2}$ | 85.13(5) | P1 | S1 | Fe1 | 103.14(8) |
| $S1^{1}$ | Fe1 | $S1^{3}$ | 85.13(5) | P1 | S1 | $Fe1^{3}$ | 103.23(8) |
| S1 | Fe1 | $S1^{3}$ | 95.51(5) | $Fe1^{4}$ | S2 | $Fe1^{6}$ | 85.15(7) |
| $S1^{2}$ | Fe1 | $S1^{3}$ | 179.05(8) | P1 | S2 | $Fe1^{6}$ | 103.24(8) |
| S1 | Fe1 | $S2^{4}$ | 179.57(5) | P1 | S2 | $Fe1^{4}$ | 103.24(8) |
| $S1^{1}$ | Fe1 | $S2^{5}$ | 179.57(5) | $S1^{7}$ | P1 | S1 | 114.41(12) |
| $S1^{1}$ | Fe1 | $S2^{4}$ | 85.12(5) | S1 | P1 | S2 | 114.05(7) |
| S1 | Fe1 | $S2^{5}$ | 85.12(5) | $S1^{7}$ | P1 | S2 | 114.05(7) |
| $S1^{2}$ | Fe1 | $S2^{4}$ | 94.44(6) | $S1^{7}$ | P1 | $P1^{8}$ | 104.39(8) |
| $S1^{3}$ | Fe1 | $S2^{4}$ | 84.91(6) | S1 | P1 | $P1^{8}$ | 104.39(8) |
| $S1^{2}$ | Fe1 | $S2^{5}$ | 84.91(6) | S2 | P1 | $P1^{8}$ | 103.88(14) |
| $S1^{3}$ | Fe1 | $S2^{5}$ | 94.44(6) | | | | |

**Table S4: Bond lengths in bulk $FePS_3$ at RT**

| Atom | Atom | Length/Å | Atom | Atom | Length/Å |
|---|---|---|---|---|---|
| Fe1 | S1 | 2.539(3) | S1 | P1 | 2.030(3) |
| Fe1 | $S1^{1}$ | 2.547(3) | S2 | $Fe1^{4}$ | 2.549(3) |
| Fe1 | $S1^{2}$ | 2.547(3) | S2 | $Fe1^{6}$ | 2.549(3) |
| Fe1 | $S1^{3}$ | 2.539(3) | S2 | P1 | 2.027(4) |
| Fe1 | $S2^{4}$ | 2.550(3) | P1 | $S1^{7}$ | 2.030(3) |
| Fe1 | $S2^{5}$ | 2.549(3) | P1 | $P1^{8}$ | 2.191(6) |
| S1 | $Fe1^{1}$ | 2.547(3) | | | |

**Table S5: Bond angles in bulk $FePS_3$ at RT**

| Atom | Atom | Atom | Angle/° | Atom | Atom | Atom | Angle/° |
|---|---|---|---|---|---|---|---|
| S1 | Fe1 | $S1^{1}$ | 94.84(13) | $S2^{4}$ | Fe1 | $S2^{5}$ | 94.61(14) |
| $S1^{1}$ | Fe1 | $S1^{2}$ | 95.57(11) | Fe1 | S1 | $Fe1^{3}$ | 84.43(11) |
| S1 | Fe1 | $S1^{2}$ | 85.09(11) | P1 | S1 | Fe1 | 103.25(14) |
| $S1^{1}$ | Fe1 | $S1^{3}$ | 85.09(11) | P1 | S1 | $Fe1^{3}$ | 103.34(11) |
| S1 | Fe1 | $S1^{3}$ | 95.57(11) | $Fe1^{4}$ | S2 | $Fe1^{6}$ | 85.39(14) |
| $S1^{2}$ | Fe1 | $S1^{3}$ | 179.03(10) | P1 | S2 | $Fe1^{6}$ | 103.29(13) |
| S1 | Fe1 | $S2^{4}$ | 179.53(7) | P1 | S2 | $Fe1^{4}$ | 103.29(13) |
| $S1^{1}$ | Fe1 | $S2^{5}$ | 179.53(7) | $S1^{7}$ | P1 | S1 | 114.46(18) |
| $S1^{1}$ | Fe1 | $S2^{4}$ | 85.28(12) | S1 | P1 | $P1^{8}$ | 104.15(10) |
| S1 | Fe1 | $S2^{5}$ | 85.28(12) | $S1^{7}$ | P1 | $P1^{8}$ | 104.15(10) |
| $S1^{2}$ | Fe1 | $S2^{4}$ | 94.44(11) | S2 | P1 | $S1^{7}$ | 114.15(11) |
| $S1^{3}$ | Fe1 | $S2^{4}$ | 84.90(11) | S2 | P1 | S1 | 114.16(11) |
| $S1^{2}$ | Fe1 | $S2^{5}$ | 84.90(11) | S2 | P1 | $P1^{8}$ | 104.04(18) |
| $S1^{3}$ | Fe1 | $S2^{5}$ | 94.44(11) | | | | |

# 2. DFT+U calculations

## 2.1. *Nearest-Neighbor exchange coupling*

To assess the impacts of the exposed distortion of the $FeS_6$ octahedra from the single XRD measurements above, nearest-neighbor (NN) spin-exchange interactions in a distorted honeycomb lattice were evaluated using Density Functional Theory (DFT). For clarity, these results were compared to $MnPS_3$ which possess an isotropic honeycomb lattice. **Table S6** exposes three distinct M–S bonds compared with $MnPS_3$. The distortion of the $[FeS_6]$ octahedra, in turn, yields inequivalent Fe–Fe distances.

**Table S6 : Anisotropies in atomic distances (MS, MM (l, l*)) and structural parameters ($\Delta_b$, $\delta$for relaxed $FePS_3$ and $MnPS_3$)**

| | a | b | δ | l | *l | MS1 | MS2 | MS3 | MS4 | MS5 | MS6 | $\|\|\Delta_b\|\|$ |
|---|---|---|---|---|---|---|---|---|---|---|---|---|
| | [Å] | | [%] | | | | | [Å] | | | | [%] |
| $FePS_3$ | 6.066 | 10.526 | 0.2 | 3.393 | 3.563 | 2.636 | 2.578 | 2.572 | 2.578 | 2.636 | 2.572 | 2.2 |
| $MnPS_3$ | 6.083 | 10.536 | 0.0 | 3.511 | 3.512 | 2.624 | 2.625 | 2.624 | 2.625 | 2.624 | 2.625 | 0.0 |

A comparison between the $\Delta_b$ values in $FePS_3$ and $MnPS_3$ (see **Table S6**), reveals unprecedented distortion of the $[FeS_6]$ octahedra, and thus conveys a symmetry breaking in the b/a plane. The distortion is determined by a parameter, δ:

$$\delta = \left(b - a\sqrt{3}\right)/a\sqrt{3} \qquad (2)$$

Notably, the magnitude of δ is larger in $FePS_3$ by ~0.2 % relative to that of $MnPS_3$. This intralayer anisotropy also impacts the NNs' spin-exchange interactions.[49] To accommodate these effects, the degrees of anisotropy in the spin-exchange couplings were assumed using a modified Heisenberg Hamiltonian, stated as:

$$H = \sum J_{1,1}\vec{S}_i \cdot \vec{S}_j + \sum J_{1,2}\vec{S}_i \cdot \vec{S}_j + \sum J_{2,1}\vec{S}_i \cdot \vec{S}_j + \sum J_{2,2}\vec{S}_i \cdot \vec{S}_j + \sum J_{3,1}\vec{S}_i \cdot \vec{S}_j + \sum J_{3,2}\vec{S}_i \cdot \vec{S}_j \quad (3)$$

where the NNs exchange interaction coefficients are given by ($J_{ij}$; i =1-3, j =1, 2). **Figure S2** depicts the $FePS_3$ honeycomb, where panel (a) labels the distinct nearest-neighbor (NN) exchange pathways ($J_{ij}$; i =1-3, j =1, 2) defined in Eq. 3, and panel (b) illustrates the a/b-zigzag magnetic configuration. The magnitudes of $J_{ij}$ were evaluated from six linear independent equations as given in the **Methods**, using DFT+U energy calculations based on the relaxed M-M distances given in **Table S6**. The $J_{ij}$ coefficients evaluated for anisotropic $FePS_3$ are listed in **Table S7**(left column), which are compared with values of isotropic $FePS_3$ (right column), as well as with an average of the anisotropic values (middle column). Interestingly, our calculations showed that isotropic $FePS_3$ values predict a false Néel magnetic ground-state with a large deviation from the zigzag-AFM magnetism reported in the literature.[50]

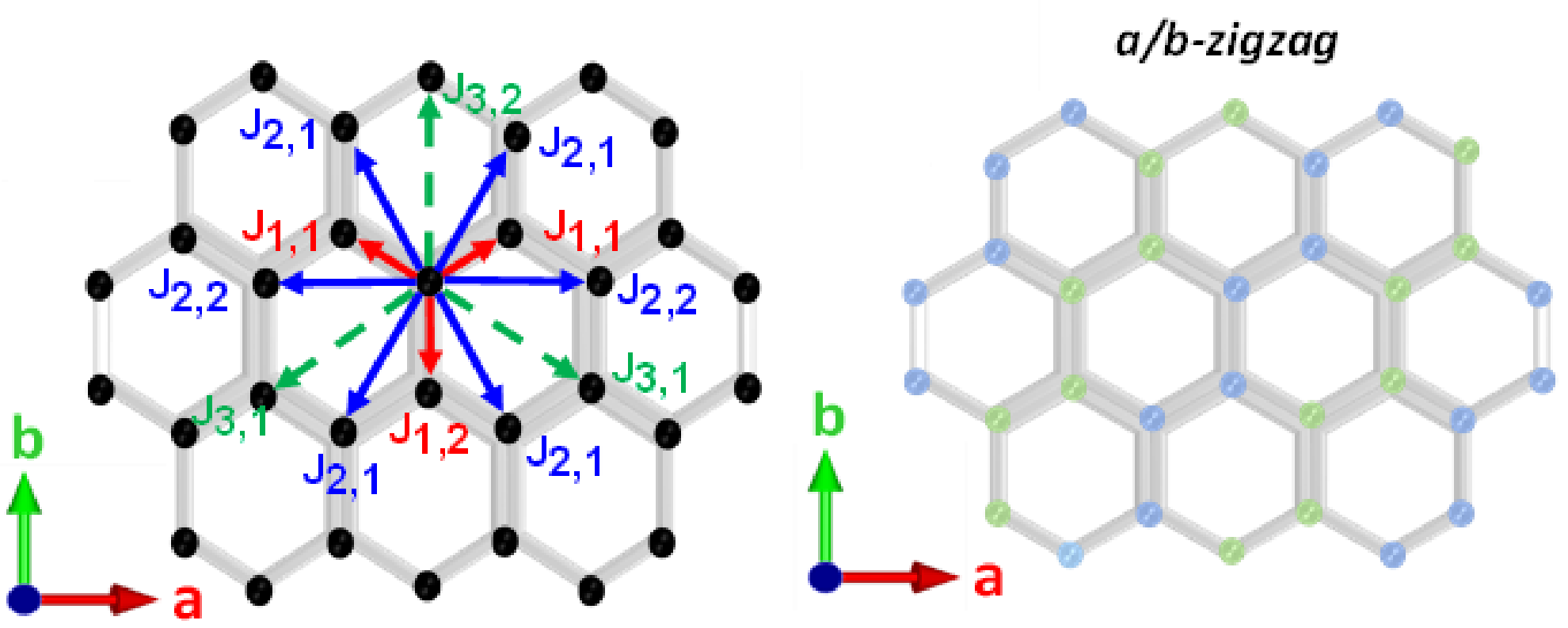

**Figure S2. FePS3 magnetic configuration. (a)** A representation of a honeycomb in FePS3 with the various nearest-neighbor spin-interactions as in Equation 3. **(b)** A representation of a honeycomb in FePS3 showing the a/b-zigzag magnetic configuration.

For anisotropic $FePS_3$, significant variance between the exchange terms is obtained. While our model accounted for structure anisotropy, other publications evaluated the isotropic spin Hamiltonian, adding spin-orbit and single-ion interactions.[16,51] Interestingly, the average exchange values agree with those of previous publications, avoiding anisotropy, emphasizing the overall importance of structural anisotropy over the spin interactions. Thus, we conclude that the non-symmetric structure dominates the magnetic behavior of $FePS_3$.

**Table S7: Comparison between spin-exchange terms in isotropic and anisotropic hexagonal network of metals in $FePS_3$**

| $FePS_3$ | | $FePS_3$ Average | | Fe in $MnPS_3$ symmetric structure | |
|---|---|---|---|---|---|
| $J_{1,1}$ | 1.42 | $\overline{J_1}$ | 2.15 | $J_1$ | 0.49 |
| $J_{1,2}$ | 3.59 | | | | |
| $J_{2,1}$ | -0.29 | $\overline{J_2}$ | -0.28 | $J_2$ | 0.17 |
| $J_{2,2}$ | -0.23 | | | | |
| $J_{3,1}$ | 0.27 | $\overline{J_3}$ | -1.13 | $J_3$ | 0.07 |
| $J_{3,2}$ | -3.94 | | | | |

**J variables for $MnPS_3$ compound:**

$$J_1 = \frac{E_{FM} - E_{N\acute{e}el} - E_{stripe} + E_{zigzag}}{8S^2}$$
$$J_2 = \frac{E_{FM} + E_{N\acute{e}el} - E_{stripe} - E_{zigzag}}{16S^2} \quad (1)$$
$$J_3 = \frac{E_{FM} - E_{N\acute{e}el} + 3E_{stripe} - 3E_{zigzag}}{24S^2}$$

## 2.2 *Electronic band structure*

Orbital-projected densities of states (DOS) were evaluated in a locally rotated coordinate system, see **Figure S3(a)**, in a similar way to M. Rassekh et. al. work on $CrI_3$.[52] The structure was rotated such that the two opposite bonds with the same length approximately align with the Cartesian $z$ axis (shown in magenta) while the two opposite bonds with different lengths (shown in red and blue) align with the $x$ and $y$ axes. Although the total density of states is invariant under this transformation (see **Figure 6** in main text and **Figure S3(b)**), the orbital-resolved decomposition depends on the chosen reference frame.

In this rotated basis, the Fe $d$ orbitals split according to the local octahedral coordination, which is distorted due to the different chemical environments of the surrounding S atoms. In particular, the in-plane bonds group into two adjacent short and two long bonds (see **Figure S3(a)**), lifting most of the orbital degeneracies. In this configuration, the $d_{xz}$ and $d_{yz}$ orbitals remain degenerate, while the remaining $d$ orbitals become non-degenerate as a consequence of the distortion.

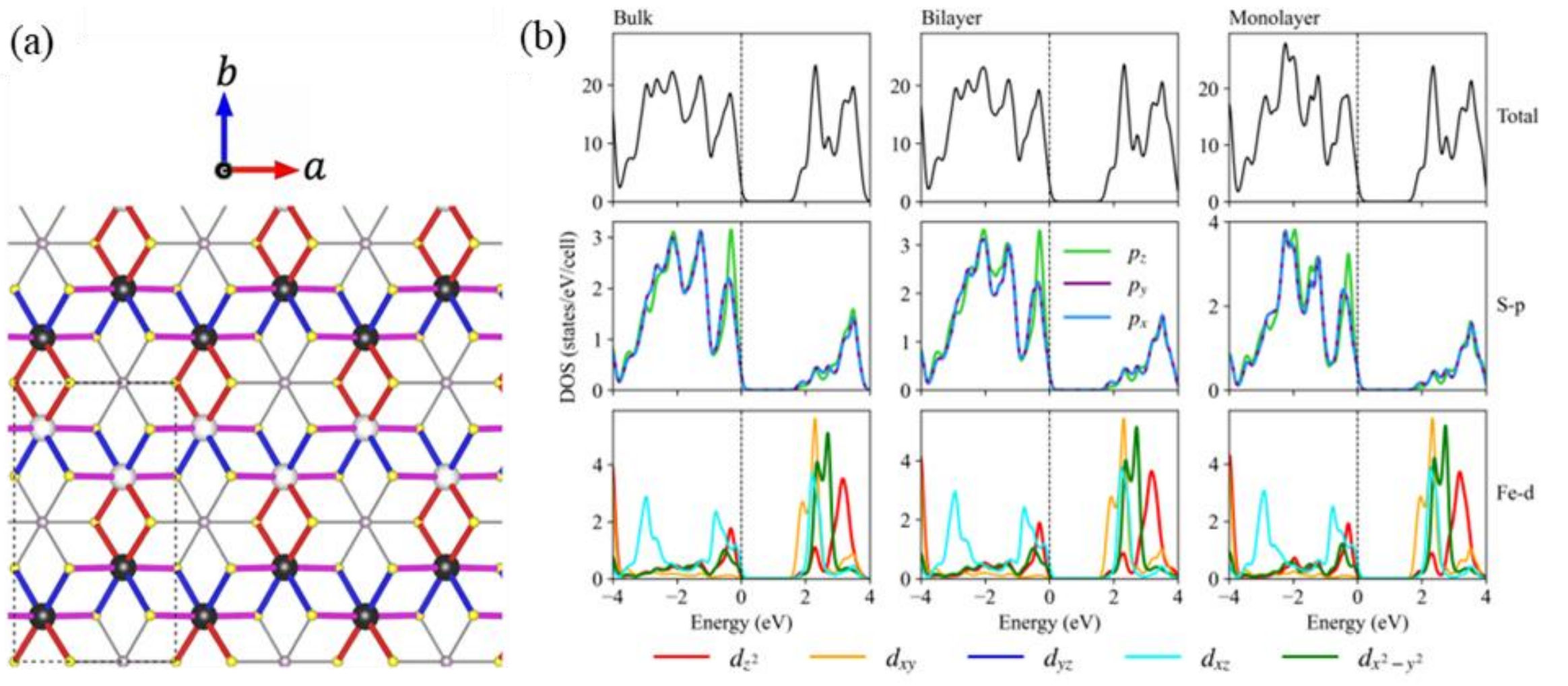

**Figure S1. Total and projected DOS of locally rotated coordinate system of $FePS_3$. (a)** Sketch of the geometry of monolayer $FePS_3$ with different Fe–S bond lengths indicated by different colors: red ~ 2.55 Å, magenta ~ 2.56 Å, and blue ~ 2.59 Å. The local octahedral coordinate system is defined such that the two opposite Fe–S bonds of equal length (magenta) are approximately aligned with the z axis, while the remaining two bonds of different lengths (red and blue) are aligned with the x and y axes. Black and white spheres indicate Fe (spin up/down), yellow spheres denote S, and rose spheres represent P. **(b)** Total and projected DOS of bulk, bilayer and monolayer $FePS_3$ calculated in the locally rotated coordinate system are given per unit cell and correspond to the sum of spin-up and spin-down components. Atomic contributions from the S and Fe components are resolved into S 3p (px, py, pz) and Fe 3d ($d_{xy}$, $d_{xz}$, $d_{yz}$, $d_{x^2-y^2}$, $d_z^2$) orbitals.

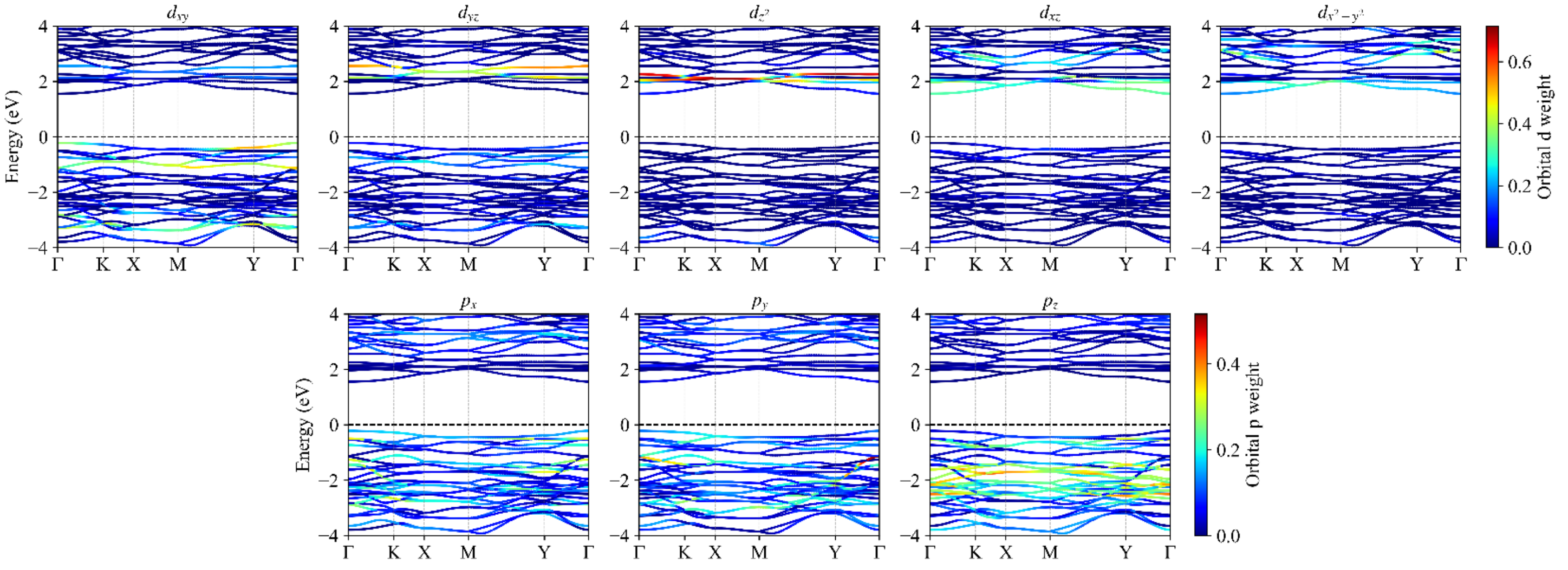

**Figure S4. Orbital-resolved band structure of monolayer $FePS_3$**. The upper panel shows the Fe 3d orbital weight ($d_{xy}, d_{yz}, d_{z^2}, d_{yz}, d_{x^2-y^2}$) and the lower panel shows the S 3p orbital weight ($p_x, p_y, p_z$). Both the valence-band maximum and the conduction-band minimum are located at the Γ point. Furthermore, the first few conduction bands show a negligible contribution from the S 3p orbitals while the valence band has a dominant in-plane character which explains the minor change of the band gap with reducing number of layers.

## 3. Second derivative Plots of $FePS_3$ emission bands

To identify whether the emission bands of $FePS_3$ consist of single or multiple components, the second derivative of each band was calculated. **Figure S5(a-d)** displays second derivative plots (red) of bands A-D (in black). The results clearly reveal that all bands are comprised of more than one component at 4 K.

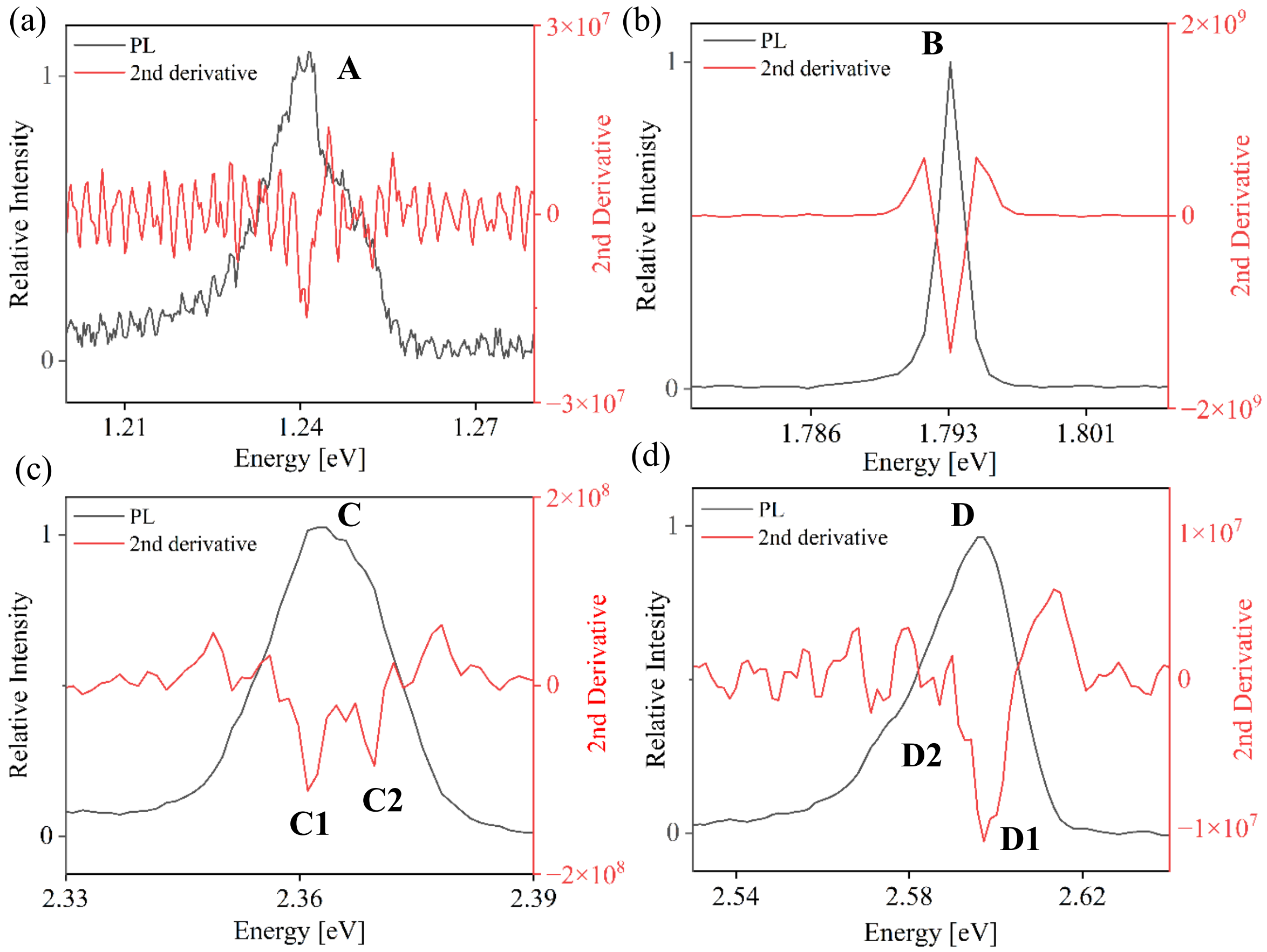

**Figure S5. Second derivative plots of emission bands of bulk $FePS_3$. (a)-(d)** μ-PL plots of bands A-D (in black) with the corresponding second derivative plots (in red), respectively.

## 4. Control Experiments on Cr:$ZnPS_3$ bulk crystal

To rule out Cr contamination as the origin of band B, a low Cr-doped control sample Cr:$ZnPS_3$, with intended composition of $Cr_{0.01}Zn_{0.99}PS_3$, was synthesized by chemical vapor transport (CVT), where elemental powders in a stoichiometric ratio, with 5% excess sulfur as the transport agent, were sealed in a quartz tube with pressure ~5 x $10^{-5}$ Torr and placed in a two-zone furnace with the substrate zone at 600 °C and the deposition zone at 550 °C. The crystals obtained were characterized by EDX. The full EDX spectrum of the crystal is shown in **Figure S6(a)** and the composition breakdown of the crystal is given in Table S8. With visible EDX peak (inset in Figure S6(a), the amount of Cr is estimated as 0.25+/-0.09 at.%, which is around 3σ, that means Cr is detectable in the control sample. **Figure S6(b, c)** shows the complete PL spectra of the material where two sharp emission bands were observed between 1.5-1.7 eV (labelled as band 1 and 2), assumed to originate from the Cr component and a broader emission band at ~2.3 eV, assumed to originate from $ZnPS_3$, as shown by Mukherjee et al.[53]

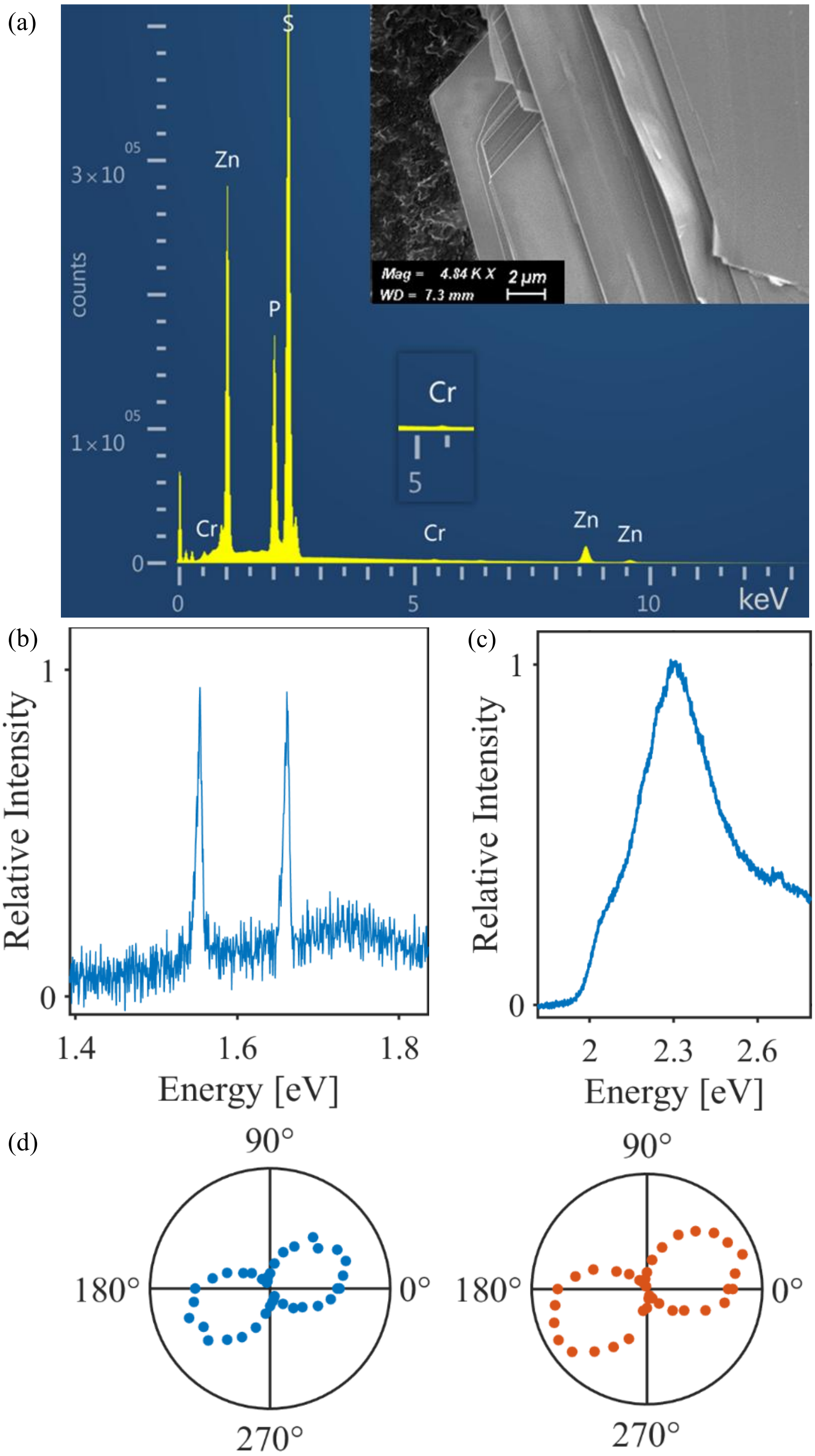

**Figure S6. Characterization and optical measurements on bulk Cr:ZnPS$_3$ single crystal. (a)** EDX spectrum of Cr:ZnPS$_3$, where the compositions of the crystal's components are shown. **Inset 1:** SEM image of Cr:ZnPS$_3$ bulk single-crystal. **Inset 2:** A region of interest presenting detectable chromium peak; **(b, c)** Emission spectra of bulk Cr:ZnPS$_3$ crystal where three emissions are revealed: **(b)** Two sharp emissions between 1.5eV-1.7eV (labelled bands 1 and 2), assumed to originate from the Cr component and **(c)** an emission at 2.3eV, assumed to originate from ZnPS$_3$; **(d)** Polar plots of bands 1 (in blue) and 2 (in orange) where a clear linear polarization behavior is observed for both emissions.

Bands 1 and 2 in the Cr:ZnPS$_3$ crystal appear sharp and narrow with a full-width-half-max (FWHM) of ~ 0.01 eV, which is comparable to the appearance of band B in the FePS$_3$ crystal shown in the main text. However, as shown in **Figure S6(d)**, the polar plots of both bands 1 and 2 from the Cr:$ZnPS_3$ sample exhibit a clear linear polarization behavior. This contrasts with band B in the $FePS_3$ samples, which did not exhibit linear polarization behavior as shown in the main text. The absence of Cr contaminants in our $FePS_3$ bulk samples and the presence of LP signature in the Cr-doped control sample, argues against Cr contamination as the origin of band B in bulk $FePS_3$.

**Table 8: Composition of bulk Cr:$ZnPS_3$ crystal from EDX measurements**

| Element | Atomic % | At. % Sigma (+/-) |
|---|---|---|
| P | 20.54 | 0.17 |
| S | 58.78 | 0.19 |
| Cr | 0.25 | 0.09 |
| Zn | 20.44 | 0.13 |
| Total: | 100.00 | |

## 5. SQUID measurements

Temperature-dependent superconducting quantum interference device (SQUID) measurements were performed on $FePS_3$ with magnetic field applied both in-plane (IP; H ∥ ab) and out-of-plane (OOP; H ⊥ ab), as shown in **Figure S7**, revealing a pronounced magnetic anisotropy across the two measurement configurations. In both, the susceptibility shows a clear cusp consistent with the onset of long-range antiferromagnetic order at $T_n$ ~120 K. However, the magnitude and curvature of $\chi(T)$ differ between IP and OOP configurations, reflecting the distinct magnetic response along and perpendicular to the layers. Below $T_n$, the stronger suppression/evolution of $\chi(T)$ in OOP orientation compared to the other is consistent with an anisotropic AFM having an easy-axis and direction-dependent susceptibility.

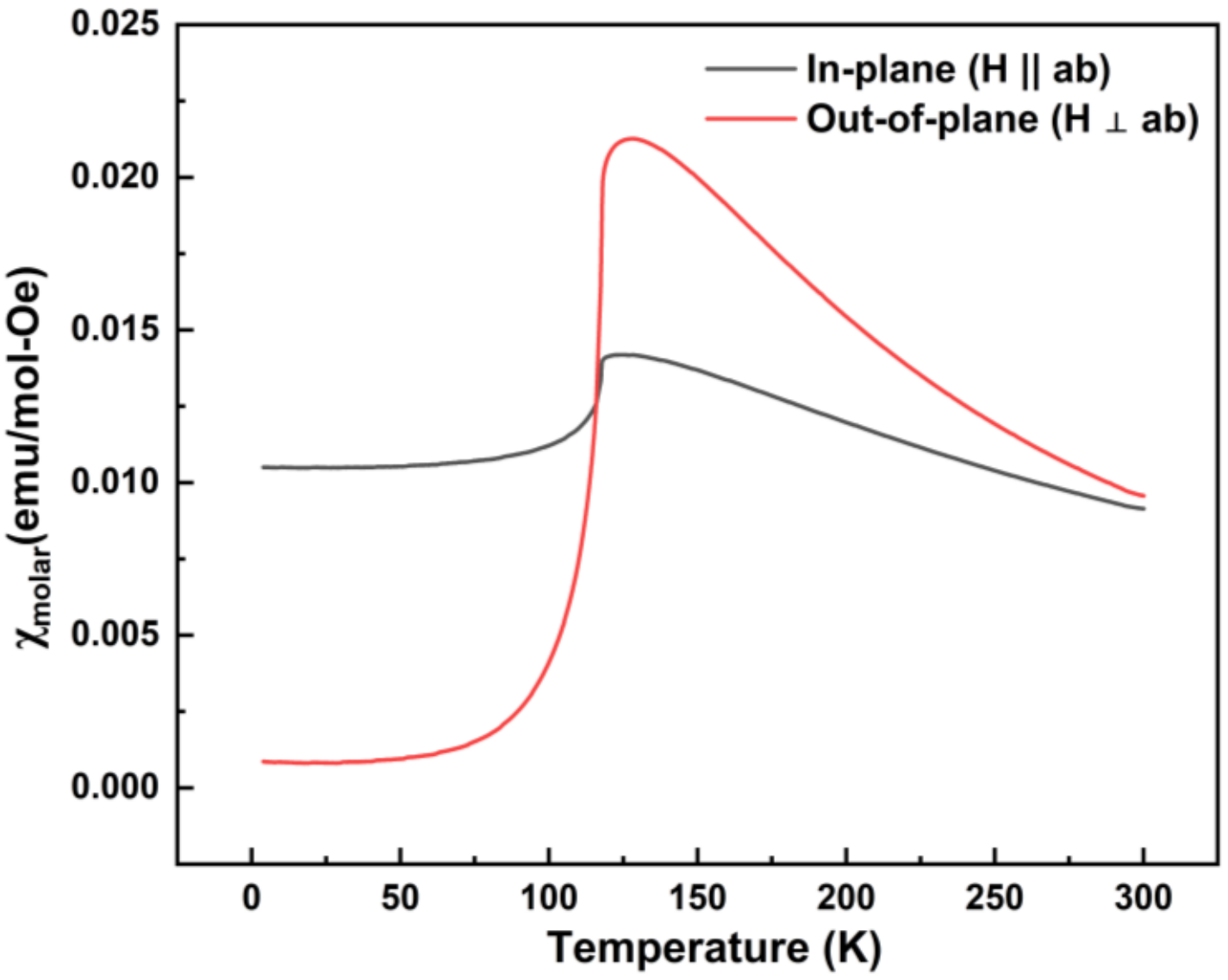

**Figure S7. Temperature-dependent magnetic susceptibility (χ) in in-plane (IP) and out-of-plane (OOP) measurement configurations,** measured under the external magnetic field of 1T.

# 6. Magneto-optical Measurements

## 6.1. *Circular polarization PL and Magneto-PL (MPL) measurements*

To amplify the modest circular polarization response observed in the emission bands B and C in both bulk and monolayer $FePS_3$ samples, PL measurements were conducted under an external magnetic field up to and including 6T. The degree of circular polarization (DCP) at varied magnetic fields was calculated as:

$$DCP(\%) = \frac{I_{\sigma^+} - I_{\sigma^-}}{I_{\sigma^+} + I_{\sigma^-}} \, x \, 100 \qquad \text{Eq. 1}$$

where $I_{\sigma+/\sigma-}$ is the intensity of the σ± circular polarized component of the MPL.

**Figure S8** shows the emission of band B under opposing circular polarizations; σ+ (in blue) and σ- (in orange) at B = 0T, 3T, and 6T in both (a) bulk and (b) monolayer $FePS_3$ samples, respectively.

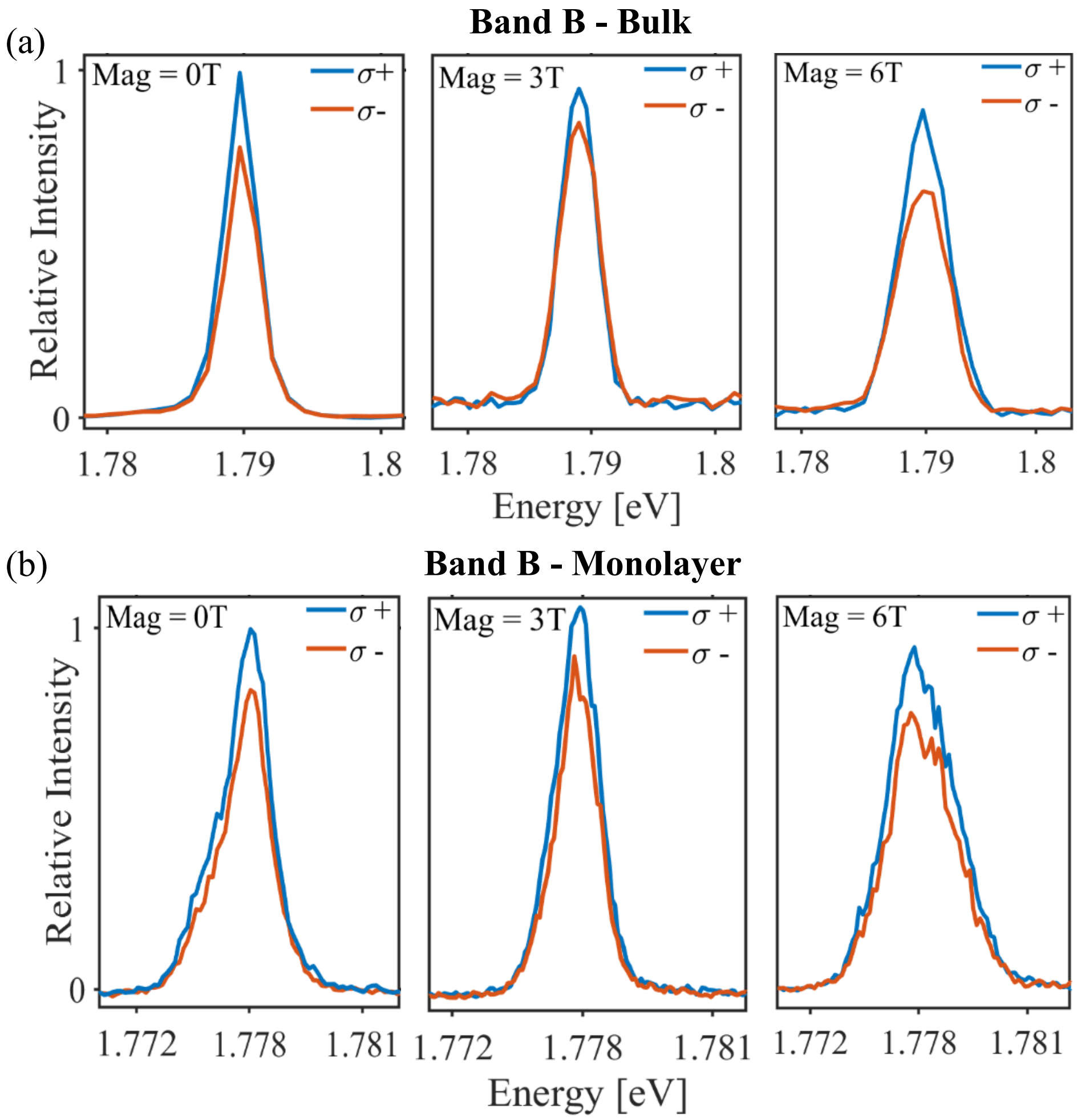

**Figure S8. Circular polarization PL and MPL measurements of emission band B in bulk and monolayer $FePS_3$**. Emission spectra of band B region under opposing circular polarizations; σ+ (in blue) and σ- (in orange) at Mag = 0T, 3T and 6T in **(a)** Bulk and **(b)** monolayer $FePS_3$ samples.

At mag = 0T, a very low DCP of ~ 10% is observed in the bulk sample (see the left panel of **Figure S8(a)**). Upon increasing the magnetic field to 3T a slight decrease in the DCP is observed (giving a DCP ~ 8%) (see middle panel of **Figure S8(a)**), followed by a return to the DCP of 10% at 6T (see right panel of **Figure S8(a)**). A similar result can be observed in the monolayer sample, see **Figure S8(b)**, whereby the DCP value remains constant at ~9%. Thus, increasing the magnetic field doesn't have a large impact on the circular polarization behavior of band B.

**Figure S9** shows the emission of band C under opposing circular polarizations; σ+ (in blue) and σ- (in orange) at Mag = 0T, 3T, and 6T, respectively in both the (a) bulk and (b) monolayer $FePS_3$ samples. In the bulk sample, at B = 0T (see the left panel of **Figure S9(a)**) both C1 and C2 exhibit a circular polarized behavior with a DCP of 20.89%, and 30.43%, respectively. Interestingly, with the increase of the magnetic field the DCP decreases substantially.

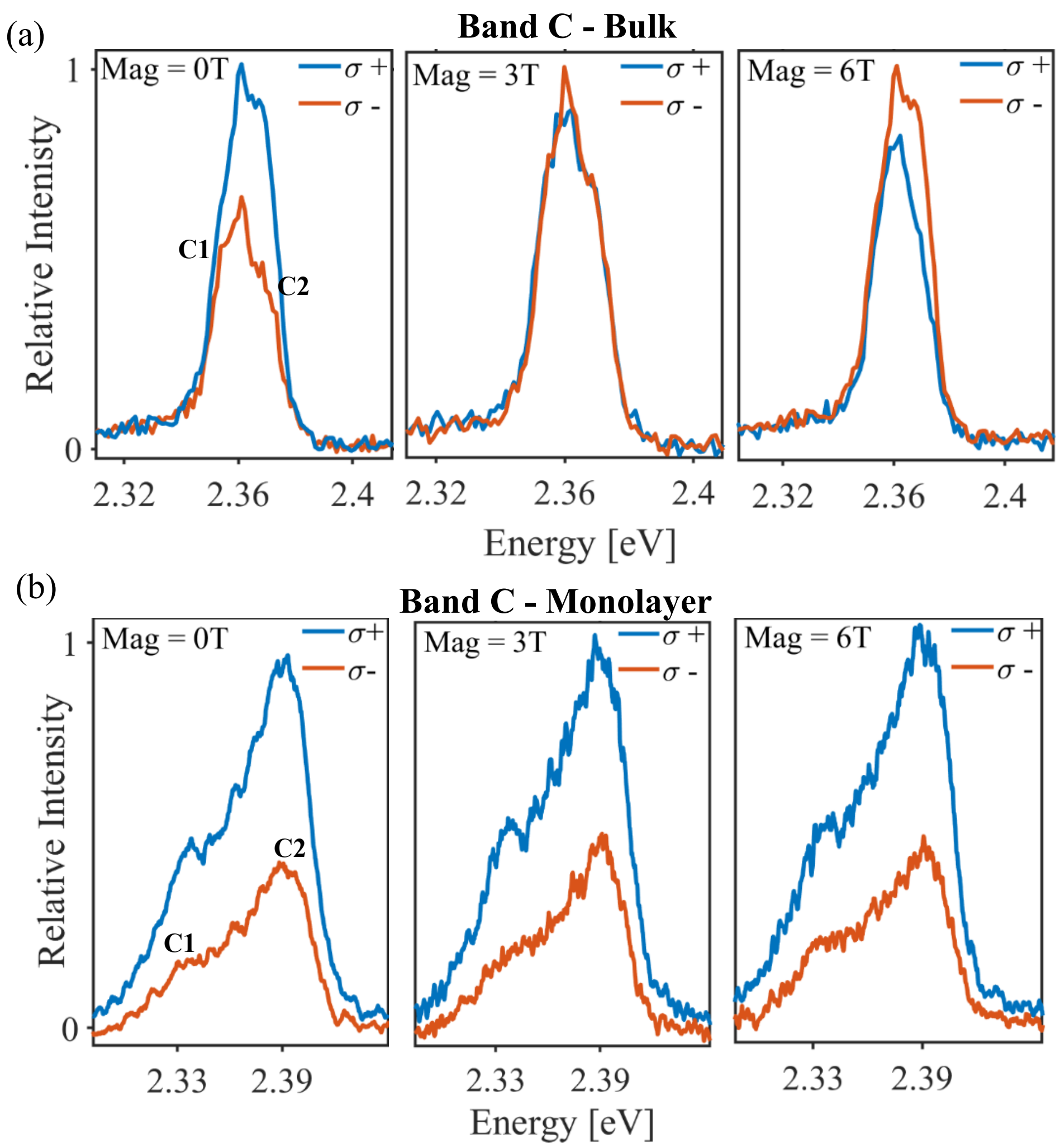

**Figure S9. Circular polarization PL and MPL measurements of emission band C in bulk and monolayer $FePS_3$.** Emission spectra of band B region under opposing circular polarizations; σ+ (in blue) and σ- (in orange) at Mag = 0T, 3T and 6T in **(a)** Bulk and **(b)** monolayer $FePS_3$ samples.

At 3T (see the middle panel of **Figure S9(a)**), the DCP drops to ~0% for both C1 and C2. At 6T (see the left panel of **Figure S9(a)**), the DCP continues to drop to ~ -5% in both C1 and C2 is observed. In contrast, in the monolayer the increase of magnetic field doesn't impact the DCP of both C1 and C2. The value remains at ~ 40% for both from 0-6T (see **Figure S9(b)**).

## 7. Temperature dependent PL measurements

To evaluate the robustness of the zigzag AFM arrangement in $FePS_3$, μPL spectra of bands B and C in the monolayer limit under increasing temperatures ranging from 4K-80K (**Figure S10**). A similar trend is observed compared to the bulk form (see **Figure 5** in main text). This is attributed to no substantial rearrangement of the CB electronic structure upon exfoliation.

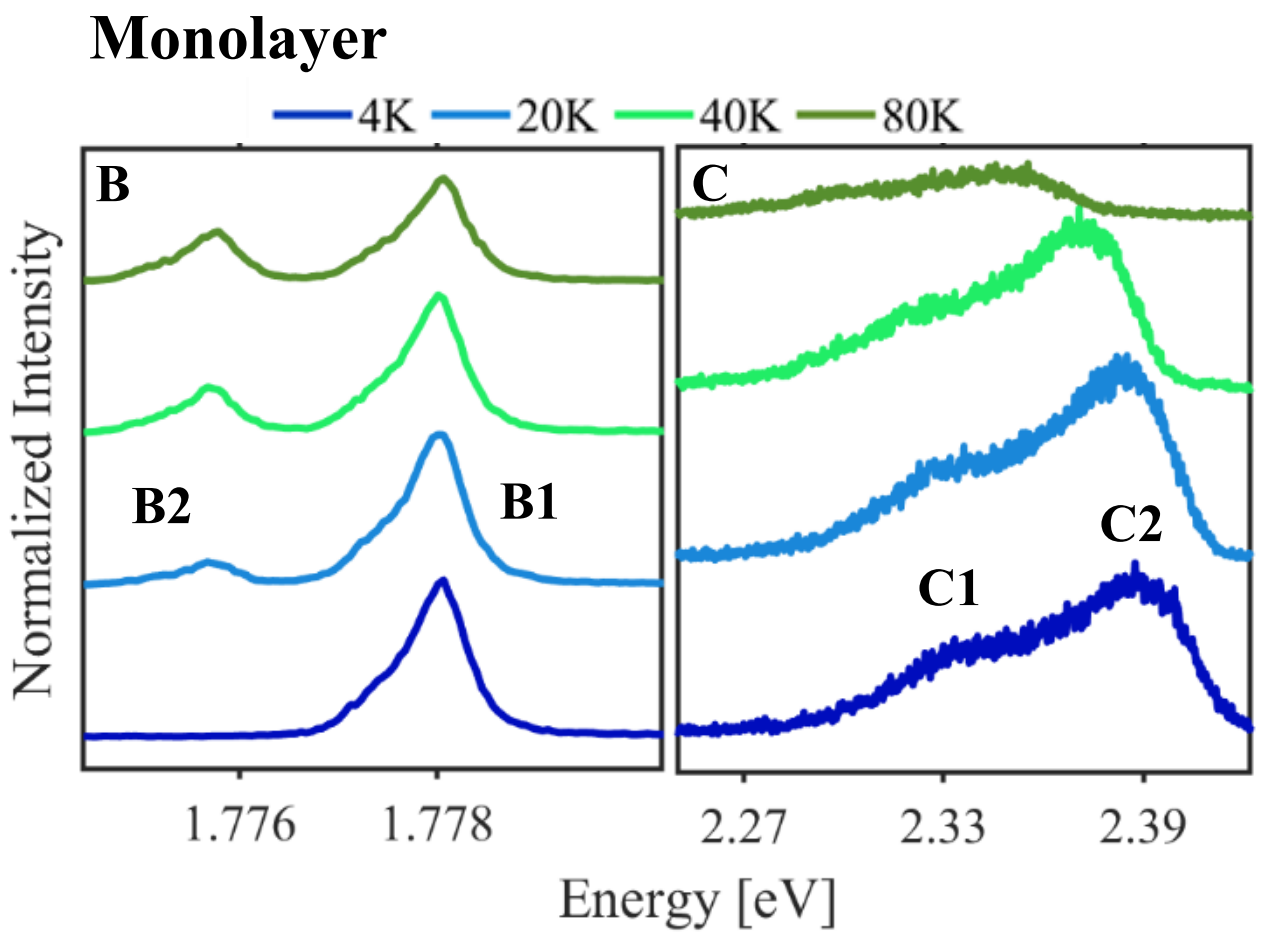

**Figure S10.** Temperature dependent PL spectra of emission bands B and C in the monolayer $FePS_3$ sample.

## References

(1) Kim, T. Y.; Park, C.H. Magnetic Anisotropy and Magnetic Ordering of Transition-Metal Phosphorus Trisulfides. *Nano Lett.* **2021**, *21*, 10114–10121.

(2) Kurosawa, K.; Saito, S.; Yamaguchi, Y. Neutron Diffraction Study on $MnPS_3$ and $FePS_3$. *J. Physical Soc. Japan* **1983**, *52*, 3919–3926.

(3) Lançon, D.; Walker, H. C.; Ressouche, E.; Ouladdiaf, B.; Rule, K. C.; McIntyre, G. J.; Hicks, T. J.; Rønnow, H. M.; Wildes, A. R. Magnetic Structure and Magnon Dynamics of the Quasi-Two-Dimensional Antiferromagnet FePS3. *Phys. Rev. B* **2016**, *94*, 214407.

(4) Olsen, T. Magnetic Anisotropy and Exchange Interactions of Two-Dimensional FePS3, NiPS3 and MnPS3 from First Principles Calculations. *J. Phys. D Appl. Phys.* **2021**, *54*, 314001.

(5) Rassekh, M.; He, J.; Farjami Shayesteh, S.; Palacios, J. J. Remarkably Enhanced Curie Temperature in Monolayer CrI3 by Hydrogen and Oxygen Adsorption: A First-Principles Calculations. *Comput. Mater. Sci.* **2020**, *183*, 109820.

(6) Mukherjee, A.; Santamaría-García, V. J.; Wlodarczyk, D.; Somakumar, A. K.; Sybilski, P.; Siebenaller, R.; Rowe, E.; Narayanan, S.; Susner, M. A.; Lozano-Sanchez, L. M.; Suchocki, A.; Palma, J. L.; Boriskina, S. V. Thermal and Dimensional Stability of Photocatalytic Material $ZnPS_3$ Under Extreme Environmental Conditions. *Adv. Electron. Mater.* **2025**, *11*.